\documentclass[aps,prd,twocolumn,nofootinbib,showpacs,superscriptaddress]{revtex4-1}
\usepackage{graphicx}
\usepackage[colorlinks=true,linkcolor=blue,citecolor=blue,urlcolor=blue]{hyperref}
\usepackage{slashed,color,amsmath,amssymb,mathrsfs}
\usepackage{multirow}
\usepackage{diagbox}
\usepackage{bbm}
\usepackage{longtable}
\usepackage{array}
\usepackage{footmisc}

\allowdisplaybreaks

\begin{document}

\title{Study of electron-positron annihilation into $K\bar{K}\pi$ within resonance chiral theory}

\author{Bing-Hai Qin}
\affiliation{School for Theoretical Physics, School of Physics and Electronics, Hunan University, Changsha 410082, People's Republic of China}
\affiliation{Hunan Provincial Key Laboratory of High-Energy Scale Physics and Applications, Hunan University, Changsha 410082, People's Republic of China}
\affiliation{School of Physics, Sun Yat-sen University, Guangzhou 510275, People’s Republic of China}

\author{Wen Qin}\email{qinwen@hunnu.edu.cn}
\affiliation{Department of Physics, Hunan Normal University, Changsha 410081, People's Republic of China}

\author{Ling-Yun Dai}\email{dailingyun@hnu.edu.cn}
\affiliation{School for Theoretical Physics, School of Physics and Electronics, Hunan University, Changsha 410082, People's Republic of China}
\affiliation{Hunan Provincial Key Laboratory of High-Energy Scale Physics and Applications, Hunan University, Changsha 410082, People's Republic of China}

\begin{abstract}
In this paper, a coherent study of the $e^+e^-$ annihilation into $K^+K^-\pi^0$, $K^0_SK^0_L\pi^0$ and $K^0_SK^\pm\pi^\mp$ is carried out within the framework of resonance chiral theory. The amplitudes are fixed by fitting to the experimental cross-section and invariant mass spectrum. With these amplitudes, one can calculate the hadronic vacuum polarization form factors of these processes. The leading order contributions of $\sigma(e^+e^- \to K\bar{K}\pi)$ to the anomalous magnetic moment of the muon, $(g-2)_\mu$, is obtained as $a_\mu^{\rm HVP,LO}(e^+e^- \to K\bar{K}\pi)=(3.178\pm0.071)\times10^{-10}$ up to $E_{\rm cm}=2.3$ GeV.
\end{abstract}

\maketitle

\section{Introduction}
Quantum Chromodynamics (QCD) is widely accepted as the basic theory of the strong interaction, of which the running coupling constant $\alpha_s$ reflects the strength of the interaction among quarks and gluons. 
Due to the asymptotic freedom nature of $\alpha_s$~\cite{Gross:1973id, Politzer:1973fx}, i.e., $\alpha_s$ decreases with increasing energy, it allows people to apply perturbative QCD (pQCD) to describe strong interactions in the high energy region, $E_{\rm cm} \geq$ 2 GeV, where $E_{\rm cm}$ is the energy in the center-of-mass frame (c.m.f.). 
However, at low energy region, $\alpha_s$ becomes too large to be taken as a perturbative parameter. 
Alternatively, one can use the effective field theory (EFT) to obtain the relevant hadronic dynamic information in the non-perturbative regime. 
As an effective theory of QCD in the low energy region, chiral perturbation theory (ChPT)~\cite{Weinberg:1978kz, Gasser:1983yg} describes the interactions between the lightest pseudoscalar mesons successfully. 
The Goldstone bosons arise due to chiral symmetry breaking, and they are treated as the degrees of freedom in EFT. Lorentz invariance, chiral symmetry, and discrete symmetries are implemented to construct the effective Lagrangians. ChPT has made significant achievements in studying the interactions between the lightest pseudoscalars composed of the lightest $u$, $d$, and $s$ valence quarks, but it is restricted to be applied at very low energies, e.g., $\pi\pi$ and $\pi K$ scattering around the threshold, as the power-counting is based on expansions of momenta.

In the middle energy region, both pQCD and ChPT fail to give an excellent description of the strong interaction. 
Further, there are many heavier resonances that appear in this energy region, such as vector $V(1^{--})$, axial-vector $A(1^{++})$ and $B(1^{+-})$. One can include these resonances as new degrees of freedom, and the chiral and discrete symmetries, i.e., the conservation of parity ($P$), charge conjugation ($C$), and hermiticity (h.c.), can still be applied to construct the interaction Lagrangians between them and the pseudoscalars. The successful theoretical tool in this regard is the resonance chiral theory (RChT)~\cite{Ecker:1988te, Ecker:1989yg, Cirigliano:2006hb, Portoles:2010yt}. It works well in the energy region where the lightest resonance states appear, e.g., ${M_\rho \leq E_{\rm cm}} \lesssim 1.5$ GeV, with $M_\rho$ the mass of $\rho(770)$. As discussed above, RChT is a theoretical tool used to address phenomenology involving resonances and pseudoscalars. It works in the energy region where the hadronic contribution to the anomalous magnetic moment of the muon is significant.

The muon anomalous magnetic moment, $a_\mu=(g-2)_\mu/2$, is one of the most precise indicators of new physics that can be both calculated reliably in the Standard Model (SM) and measured in experiments with very high statistics~\cite{Jegerlehner:2017gek}. Therefore, the deep understanding of $a_\mu$ may give a definitive answer to the question of whether there exists new physics beyond the SM or not. 
In 2021, the Fermi National Accelerator Laboratory (FNAL) published an accurate experimental result on the muon anomalous magnetic moment, $a_\mu({\rm Exp})=116\,592\,040(54) \times 10^{-11} ~ (0.46 ~ {\rm ppm})$~\cite{Muong-2:2021ojo}. After combining the result measured by the Brookhaven National Laboratory (BNL)~\cite{Muong-2:2006rrc}, it deviated from the theoretical prediction of SM by 4.2 $\sigma$. See Ref.~\cite{Aoyama:2020ynm} and references therein, e.g., Refs.~\cite{Aoyama:2012wk, Aoyama:2019ryr, Jackiw:1972jz, Knecht:2002hr, Czarnecki:2002nt, Gnendiger:2013pva, Colangelo:2018mtw, Keshavarzi:2018mgv, Davier:2019can, Keshavarzi:2019abf, Prades:2009tw, Guevara:2018rhj, Colangelo:2019uex, Danilkin:2019opj}. 
This reveals a possible new physics signal and draws extensive attention to particle physics. 
Very recently, the FNAL released their latest result in August 2023, $a_\mu({\rm Exp})=116\,592\,055(24) \times 10^{-11} ~ (0.20 ~ {\rm ppm})$~\cite{Muong-2:2023cdq}, leading to an average value of experimental measurements as $a^{\rm avg}_\mu({\rm Exp})=116\,592\,059(22) \times 10^{-11} ~ (0.19 ~ {\rm ppm})$. 
Now the discrepancy between the theoretical prediction of SM and the experimental measurement reaches $\Delta a_\mu=a^{\rm avg}_\mu({\rm Exp})-a_\mu({\rm SM})=(249 \pm 48) \times 10^{-11}$, with a significance of 5.1 $\sigma$. 
Nevertheless, the theoretical predictions of different models have apparent discrepancies. Different from the data-driven method~\cite{Colangelo:2018mtw, Keshavarzi:2018mgv, Davier:2019can, Keshavarzi:2019abf}, the lattice QCD gives a much larger estimation of the hadronic vacuum polarization (HVP)~\cite{Borsanyi:2020mff}, resulting in a much closer theoretical prediction to the experimental value, i.e., a much smaller significance. 
There are some other results from different lattice groups, e.g., Refs.~\cite{Ce:2022kxy, ExtendedTwistedMass:2022jpw, FermilabLatticeHPQCD:2023jof, RBC:2023pvn} which consider Euclidean window quantities and not the total HVP, displaying a roughly 3.6-3.9 $\sigma$ discrepancy from that of the data-driven method. 
Besides, it is worth mentioning that the latest measurement of the cross-section of $e^+e^- \to \pi\pi$ by CMD-3~\cite{CMD-3:2023alj, CMD-3:2023rfe} is larger than the previous measurements, e.g., Refs.~\cite{Achasov:2006vp, CMD-2:2006gxt, BaBar:2012bdw, KLOE-2:2017fda, BESIII:2015equ, Xiao:2017dqv}, implying a larger HVP contribution (to the theoretical $a_\mu$) which is close to the results of the lattice QCD.

In the theoretical predictions, the most significant uncertainties come from the HVP and the hadronic light-by-light (HLBL) scatterings, where the former has even larger uncertainties than the latter. The essential error source of HVP is from the processes of electron-positron annihilation into the lightest pseudoscalar mesons. 
In our previous studies~\cite{Qin:2020udp, Wang:2023njt}, we have studied the $e^+e^-$ annihilation processes with some of the two and three pseudoscalar final states, i.e., $\pi\pi$, $K\bar{K}$, $\pi\gamma$, $\eta\gamma$ and $\pi\pi\pi$, $\pi\pi\eta$ within the framework of RChT, and their contributions to the $(g-2)_\mu$ are given. 
In the present analysis, we will continue to analyze the processes of $e^+e^- \to K\bar{K}\pi$ with a similar theoretical framework as before. 
Unlike the case of $e^+e^- \to \pi\pi\pi$, 
the tensor mesons appear as the intermediate states in the process of $e^+e^- \to K\bar{K}\pi$ and play an important role in the interactions with pseudoscalars, vectors, and photons. This needs further systematical study within RChT~\footnote{For previous works that focus on the interactions of tensors, we refer to Refs.~\cite{Bellucci:1994eb, Toublan:1995bk, Chow:1997sg, Giacosa:2005bw, Ecker:2007us, Kubis:2015sga}. Recently, Ref.~\cite{Chen:2023ybr} studies the properties of the lightest tensor nonet with RChT.}. 
Other than the interaction Lagrangians of one tensor coupling with two pseudoscalars (TPP)~\cite{Toublan:1995bk, Ecker:2007us}, we build all the other Lagrangians associated with one tensor, e.g., tensor-electromagnetic current-pseudoscalar (TJP), tensor-vector-pseudoscalar (TVP). They are constructed in the framework of RChT, taking into account the Lorentz invariance, chiral symmetry, and discrete symmetries.

On the experimental side, the process $e^+e^- \to K\bar{K}\pi$ has been measured a few times by different collaborations, such as DM1, DM2, BABAR, SND, CMD-3, and BESIII from 1982 to 2024~\cite{Mane:1982si, Bisello:1991kd, BaBar:2007ceh, Solodov:2016klh, BaBar:2017nrz, Achasov:2017vaq, Semenov:2019zzi, Uskov:2020xna, SND:2020qmb, BESIII:2022wxz, BESIII:2023xac}. 
Especially there are also measurements about two body invariant mass spectra~\cite{Achasov:2017vaq, SND:2020qmb}, which contain the Dalitz plot information and can be somewhat helpful for refining the amplitudes~\cite{Dai:2014zta, Dai:2014lza, Yao:2020bxx}. 
These datasets, together with the theoretical tools of RChT, provide an appropriate way to refine the analysis of the processes of $e^+e^- \to K\bar{K}\pi$.

The remaining parts of the paper are organized as follows. In section~\ref{sec:2}, we will briefly introduce the theoretical framework based on RChT and construct all the required interaction Lagrangians. Then, we derive the form factors for $K\bar{K}\pi$ and discuss the high energy behavior constraints. In section~\ref{sec:4}, we will show our numerical results and obtain the HVP contributions of $\sigma(e^+e^- \to K\bar{K}\pi)$. Finally, we summarize the conclusions.

\section{Theoretical framework} \label{sec:2}
\subsection{Effective Lagrangians}
As mentioned above, at very low energies, ChPT describes well the interactions of pseudoscalar Goldstone bosons generated by the spontaneous breaking of chiral symmetry. These pseudoscalars ($\pi, K, \eta$) can be filled in an octet field $\Phi$, which is realized nonlinearly by the unitary matrix in the flavor space
\begin{eqnarray}
    u(\Phi)={\rm exp} \left\{ \frac{i}{\sqrt{2}F}\Phi \right\}~,
\end{eqnarray}
with
\begin{eqnarray}
    \Phi=
    \left(
    \begin{array}{c c c}
    \frac{\pi^0}{\sqrt{2}}+\frac{\eta_8}{\sqrt{6}}     &     \pi^+          &     K^+ \\
    \pi^-     &        -\frac{\pi^0}{\sqrt{2}}+\frac{\eta_8}{\sqrt{6}}      &     K^0 \\
    K^-       &         \bar{K}^0                      &    -\frac{2\eta_8}{\sqrt{6}} \\
    \end{array} \nonumber
    \right)~,
\end{eqnarray}
where $F$ is the pion decay constant, and its value is taken as $92.2$ MeV~\cite{ParticleDataGroup:2022pth}. 
The physical states $\eta$ and $\eta'$ are composed of a mixture of the octet $\eta_8$ and the singlet $\eta_0$ through the mixing angle $\theta_P$:
\begin{eqnarray}
    \binom{\eta_8}{\eta_0} = \binom{\cos\theta_P  ~ ~ ~ ~  \sin\theta_P}{-\sin\theta_P  ~ ~  \cos\theta_P} \binom{\eta}{\eta'}~.
\end{eqnarray}
In this analysis, $K_S^0$ and $K_L^0$ are in final states and they can be written in terms of $K^0-\bar{K}^0$ mixture, i.e., $|K_S^0\rangle = 1/\sqrt{2}~(|K^0\rangle + |\bar{K}^0\rangle)$ and $|K_L^0\rangle = 1/\sqrt{2}~(|K^0\rangle - |\bar{K}^0\rangle)$ with the assumption of conservation of $CP$ transformations.

In the intermediate energy regions, there appears a large number of resonances, e.g., vectors $V(\rho, \omega, \phi, K^\ast)$ and tensors $T(a_2, f_2, f_2', K_2^\ast)$. ChPT encounters difficulties when it comes to the energy region involving the interactions of these resonances, and RChT can be applied to expand the working energy region of ChPT, where the resonances are filled in the octets and singlets as
\begin{eqnarray}
    R = \sum_{i=1}^8 \frac{\lambda_i}{\sqrt{2}}R_i + \frac{R_0}{\sqrt{3}}\mathbbm{1}~,
\end{eqnarray}
where $R=V,T$ denotes vector and tensor resonances, respectively. The vector mesons are described by the anti-symmetric tensor field~\cite{Dai:2013joa}, which can be filled in an explicit matrix form as
\begin{eqnarray}
    V_{\mu\nu}\!=\!
    \left(
    \begin{array}{c c c}
    \frac{\rho^0}{\sqrt{2}}+\frac{\omega_8}{\sqrt{6}}+\frac{\omega_0}{\sqrt{3}}     &      \rho^+    &     K^{\ast +} \\
    \rho^-     &     -\frac{\rho^0}{\sqrt{2}}+\frac{\omega_8}{\sqrt{6}}+\frac{\omega_0}{\sqrt{3}}    &     K^{\ast 0} \\
    K^{\ast -} &      \bar{K}^{\ast 0}                    &     -\frac{2\omega_8}{\sqrt{6}}+\frac{\omega_0}{\sqrt{3}} \\
    \end{array}
    \!\right)_{\mu\nu}\!\!\!, \nonumber
\end{eqnarray}
and the tensor mesons are described by the symmetric tensor field~\cite{Ecker:2007us}
\begin{eqnarray}
    T_{\mu\nu}\!=\!
    \left(
    \begin{array}{c c c}
    \frac{a_2^0}{\sqrt{2}}+\frac{f_2^8}{\sqrt{6}}+\frac{f_2^0}{\sqrt{3}}     &     a_2^+         &     K_2^{\ast +} \\
    a_2^-        &     -\frac{a_2^0}{\sqrt{2}}+\frac{f_2^8}{\sqrt{6}}+\frac{f_2^0}{\sqrt{3}}     &     K_2^{\ast 0} \\
    K_2^{\ast -} &      \bar{K}_2^{\ast 0}             &            -\frac{2f_2^8}{\sqrt{6}}+\frac{f_2^0}{\sqrt{3}} \\
    \end{array}
    \!\right)_{\mu\nu}\!\!\!, \nonumber
\end{eqnarray}
where $V_{\mu\nu}=-V_{\nu\mu}$ and $T_{\mu\nu}=T_{\nu\mu}$. For simplicity, all states are considered to be ideal mixing if not specified. For example, the $\rho^0-\omega$ mixing is ignored. The octet $\omega_8$ and singlet $\omega_0$ of the vectors can be written as a linear combination of physical $\omega$ and $\phi$ resonances,
\begin{eqnarray}
    \omega_8 = \sqrt{\frac{1}{3}}~\omega + \sqrt{\frac{2}{3}}~\phi~, ~ ~ ~ \omega_0 = \sqrt{\frac{2}{3}}~\omega - \sqrt{\frac{1}{3}}~\phi~. \nonumber
\end{eqnarray}
The mixing mechanism of the tensors is similar to the vectors. One has
\begin{eqnarray}
    f_2^8 = \sqrt{\frac{1}{3}}~f_2 + \sqrt{\frac{2}{3}}~f_2'~, ~ ~ ~ f_2^0 = \sqrt{\frac{2}{3}}~f_2 - \sqrt{\frac{1}{3}}~f_2'~. \nonumber
\end{eqnarray}

The interaction Lagrangians can be divided into two parts: one is the interaction between the pseudoscalars themselves, $\mathcal{L}^{\rm GB}$, where $GB$ represents the Goldstone bosons, while the other part contains at least one resonance, $\mathcal{L}^{\rm R}$. 
According to the chiral counting, the Lagrangians up to $\mathcal{O}(p^4)$~\cite{Cirigliano:2006hb} will be considered in this analysis. 
The total interaction Lagrangians of RChT is
\begin{eqnarray}
    \mathcal{L}_{\rm RChT} = \mathcal{L}^{\rm GB} + \mathcal{L}^{\rm R}_{\rm kin} + \mathcal{L}^{\rm R}_{\rm int}~,
\end{eqnarray}
where $\mathcal{L}^{\rm R}_{\rm kin}$ stands for the kinetic term of the resonances, and $\mathcal{L}^{\rm R}_{\rm int}$ is for the interaction term. Further, the interaction term is
\begin{eqnarray}
    \mathcal{L}^{\rm R}_{\rm int} &=& \mathcal{L}^{\rm R}_{(2)} + \mathcal{L}^{\rm R}_{(4)} + \mathcal{L}^{\rm RR}_{(2)} \nonumber \\
    &=& \mathcal{L}^{\rm V}_{(2)} + \mathcal{L}^{\rm V}_{(4)} + \mathcal{L}^{\rm VV}_{(2)} + \mathcal{L}^{\rm T}_{(2)} + \mathcal{L}^{\rm T}_{(4)} + \mathcal{L}^{\rm TV}_{(2)}~,
\end{eqnarray}
where the number in the bracket of the subscripts represents the order of the chiral counting, and $R$, $RR$ in the superscripts denote the interactions involving one or two resonances, respectively.

For the process of $e^+e^-$ annihilation into $K\bar{K}\pi$, the leading order contribution to $\mathcal{L}^{\rm GB}$ is from the Wess-Zumino-Witten (WZW) anomaly~\cite{Wess:1971yu, Witten:1983tw}, at $\mathcal{O}(p^4)$, which is of odd-intrinsic-parity~\cite{Dai:2013joa},
\begin{eqnarray}
    \mathcal{L}^{\rm GB}_{(4)} = i \frac{N_C\sqrt{2}}{12\pi^2F^3} \varepsilon_{\mu\nu\rho\sigma} \langle \partial^\mu\Phi \partial^\nu\Phi \partial^\rho\Phi v^\sigma \rangle~,
\end{eqnarray}
where the operator $v^\sigma$ is the external vector current. The more explicit interaction Lagrangians after expansion are
\begin{eqnarray}
    \mathcal{L}^{\rm WZW}_{(4)} &=& -ie \frac{N_C}{12\pi^2F^3}\varepsilon_{\mu\nu\rho\sigma}\partial^\mu K^+ \partial^\nu K^- \partial^\rho \pi^0 A^\sigma \nonumber \\
    && -ie\frac{N_C}{12\pi^2F^3}\varepsilon_{\mu\nu\rho\sigma}\partial^\mu K^0 \partial^\nu \bar{K}^0 \partial^\rho \pi^0 A^\sigma~.
\end{eqnarray}
The kinetic Lagrangians for the lightest vector mesons are given as
\begin{eqnarray}
    \mathcal{L}^{\rm V}_{\rm kin} = -\frac{1}{2} \langle \nabla^\lambda V_{\lambda\mu} \nabla_\nu V^{\nu\mu} \rangle + \frac{1}{4} M_V^2 \langle V_{\mu\nu}V^{\mu\nu} \rangle~.
\end{eqnarray}
The explicit forms of the interaction Lagrangians involved with one vector meson are as follows~\cite{Ruiz-Femenia:2003jdx, Dumm:2009kj}
\begin{eqnarray}
    \mathcal{L}^{\rm V}_{(2)} = \frac{F_V}{2\sqrt{2}} \langle V_{\mu\nu} f^{\mu\nu}_+ \rangle + i \frac{G_V}{\sqrt{2}} \langle V_{\mu\nu} u^\mu u^\nu \rangle~,
\end{eqnarray}
\begin{eqnarray}
    \mathcal{L}^{\rm V}_{(4)} = \sum^7_{j=1} \frac{c_j}{M_V} \mathcal{O}^j_{\rm VJP} + \sum^5_{j=1} \frac{g_j}{M_V} \mathcal{O}^j_{\rm VPPP}~,
\end{eqnarray}
with
\begin{eqnarray}
    \mathcal{O}^1_{\rm VJP} &=&  \varepsilon_{\mu\nu\rho\sigma} \langle \{ V^{\mu\nu},f^{\rho\alpha}_+ \} \nabla_\alpha u^\sigma \rangle~, \nonumber \\
    \mathcal{O}^2_{\rm VJP} &=&  \varepsilon_{\mu\nu\rho\sigma} \langle \{ V^{\mu\alpha},f^{\rho\sigma}_+ \} \nabla_\alpha u^\nu \rangle~, \nonumber \\
    \mathcal{O}^3_{\rm VJP} &=& i\varepsilon_{\mu\nu\rho\sigma} \langle \{ V^{\mu\nu},f^{\rho\sigma}_+ \} \chi_- \rangle~, \nonumber \\
    \mathcal{O}^4_{\rm VJP} &=& i\varepsilon_{\mu\nu\rho\sigma} \langle V^{\mu\nu} [ f^{\rho\sigma}_-,\chi_+ ] \rangle~, \nonumber \\
    \mathcal{O}^5_{\rm VJP} &=&  \varepsilon_{\mu\nu\rho\sigma} \langle \{ \nabla_\alpha V^{\mu\nu},f^{\rho\alpha}_+ \} u^\sigma \rangle~, \nonumber \\
    \mathcal{O}^6_{\rm VJP} &=&  \varepsilon_{\mu\nu\rho\sigma} \langle \{ \nabla_\alpha V^{\mu\alpha},f^{\rho\sigma}_+ \} u^\nu \rangle~, \nonumber \\
    \mathcal{O}^7_{\rm VJP} &=&  \varepsilon_{\mu\nu\rho\sigma} \langle \{ \nabla^\sigma V^{\mu\nu},f^{\rho\alpha}_+ \} u_\alpha \rangle~, \\[3mm]
\mathcal{O}^1_{\rm VPPP} &=& i\varepsilon_{\mu\nu\alpha\beta} \langle V^{\mu\nu} ( h^{\alpha\gamma}u_\gamma u^\beta-u^\beta u_\gamma h^{\alpha\gamma} ) \rangle~, \nonumber \\
\mathcal{O}^2_{\rm VPPP} &=& i\varepsilon_{\mu\nu\alpha\beta} \langle V^{\mu\nu} ( h^{\alpha\gamma}u^\beta u_\gamma-u_\gamma u^\beta h^{\alpha\gamma} ) \rangle~, \nonumber \\
\mathcal{O}^3_{\rm VPPP} &=& i\varepsilon_{\mu\nu\alpha\beta} \langle V^{\mu\nu} ( u_\gamma h^{\alpha\gamma} u^\beta-u^\beta h^{\alpha\gamma} u_\gamma ) \rangle~, \nonumber \\
\mathcal{O}^4_{\rm VPPP} &=&  \varepsilon_{\mu\nu\alpha\beta} \langle \{ V^{\mu\nu},u^\alpha u^\beta \} \chi_- \rangle~, \nonumber \\
\mathcal{O}^5_{\rm VPPP} &=&  \varepsilon_{\mu\nu\alpha\beta} \langle u^\alpha V^{\mu\nu} u^\beta \chi_- \rangle~.
\end{eqnarray}
The interaction Lagrangians with two vector mesons is given as~\cite{Ruiz-Femenia:2003jdx, Dumm:2009kj}
\begin{eqnarray}
    \mathcal{L}_{(2)}^{\rm VV} = \sum_{j=1}^{4} d_j \mathcal{O}_{\rm VVP}^{j}~,
\end{eqnarray}
with the explicit forms
\begin{eqnarray}
    \mathcal{O}^1_{\rm VVP} &=&  \varepsilon_{\mu\nu\rho\sigma} \langle \{ V^{\mu\nu},V^{\rho\alpha} \} \nabla_\alpha u^\sigma \rangle~, \nonumber \\
    \mathcal{O}^2_{\rm VVP} &=& i\varepsilon_{\mu\nu\rho\sigma} \langle \{ V^{\mu\nu},V^{\rho\sigma} \} \chi_- \rangle~, \nonumber \\
    \mathcal{O}^3_{\rm VVP} &=&  \varepsilon_{\mu\nu\rho\sigma} \langle \{ \nabla_\alpha V^{\mu\nu},V^{\rho\alpha} \} u^\sigma \rangle~, \nonumber \\
    \mathcal{O}^4_{\rm VVP} &=&  \varepsilon_{\mu\nu\rho\sigma} \langle \{ \nabla^\sigma V^{\mu\nu},V^{\rho\alpha} \} u_\alpha \rangle~.
\end{eqnarray}

The Lagrangians involved with tensor mesons are as follows: the kinetic term is given as~\cite{Bellucci:1994eb, Ecker:2007us}
\begin{eqnarray}
    \mathcal{L}^{\rm T}_{\rm kin} = -\frac{1}{2} \langle T_{\mu\nu}D_T^{\mu\nu,\rho\sigma}T_{\rho\sigma} \rangle~,
\end{eqnarray}
where one has
\begin{eqnarray}
D_T^{\mu\nu,\rho\sigma} \!\!&=&\!\! (\Box \!+\! M_T^2) \bigg[ \frac{1}{2}(g^{\mu\rho}g^{\nu\sigma} + g^{\mu\sigma}g^{\nu\rho}) - g^{\mu\nu}g^{\rho\sigma} \bigg] \nonumber \\
&& +g^{\rho\sigma}\partial^\mu\partial^\nu \!+\! g^{\mu\nu}\partial^\rho\partial^\sigma 
\!-\! \frac{1}{2} \big( g^{\nu\sigma}\partial^\mu\partial^\rho \!+\! g^{\rho\nu}\partial^\mu\partial^\sigma \nonumber \\
&& +g^{\mu\sigma}\partial^\rho\partial^\nu + g^{\rho\mu}\partial^\sigma\partial^\nu \big)~, \nonumber
\end{eqnarray}
with $\Box$ the d'Alembert operator. The tensor Feynman propagator $G_{\mu\nu,\rho\sigma}^T(x)$ and polarization tensor $\varepsilon_{\mu\nu}$ are given in the Appendix~\ref{App:1}. 
The interaction term of $\mathcal{L}^{\rm T}_{(2)}$ of the lowest chiral counting is~\cite{Bellucci:1994eb, Ecker:2007us}
\begin{eqnarray}
    \mathcal{L}^{\rm T}_{(2)} = \langle T_{\mu\nu} J^{\mu\nu}_T \rangle~.
\end{eqnarray}
Here, the current $J^{\mu\nu}_T=J^{\nu\mu}_T$ is also symmetric, and it consists of two parts at $\mathcal{O}(p^2)$~\cite{Ecker:2007us}:
\begin{eqnarray}
    J^{\mu\nu}_T = g_T \{u^\mu,u^\nu\} + g^{\mu\nu}(\beta u^\mu u_\mu + \gamma \chi_+)~. \nonumber
\end{eqnarray}
Notice that in the calculation of the interactions between one tensor and two pseudoscalars, i.e., TPP vertex, the second part of $J^{\mu\nu}_T$ has no contributions at leading order, and one can ignore the relevant terms with couplings $\beta$ or $\gamma$. Hence, the lowest order Lagrangian of TPP is
\begin{eqnarray}
    \mathcal{L}^{\rm T}_{(2)} \supset \mathcal{L}_{\rm TPP}^{(2)} = g_T \langle T_{\mu\nu} \{u^\mu,u^\nu\} \rangle~.
\end{eqnarray}
The other interaction Lagrangians relevant to the tensors are of odd-intrinsic parity, e.g., $\mathcal{L}_{\rm TJP}^{(4)}$ and $\mathcal{L}_{\rm TVP}^{(2)}$. These Lagrangians had not been given in RChT before~\cite{Kubis:2015sga}. Here, we construct them concerning the chiral symmetry, discrete symmetries, $P$ and $C$, and hermiticity. Details can be found in Appendix~\ref{App:2}. They are at order $\mathcal{O}(p^4)$ for TJP vertexes and $\mathcal{O}(p^2)$ for TVP ones,
\begin{eqnarray}
    \mathcal{L}^{\rm T}_{(4)}  & \supset & \mathcal{L}_{\rm TJP}^{(4)} = \sum_{j=1}^3 c_j^T \mathcal{O}_{\rm TJP}^j~, \\
    \mathcal{L}^{\rm TV}_{(2)} & \supset & \mathcal{L}_{\rm TVP}^{(2)} = \sum_{j=1}^3 d_j^T \mathcal{O}_{\rm TVP}^j~,
\end{eqnarray}
with
\begin{eqnarray}
\mathcal{O}_{\rm TJP}^1 &=& i\varepsilon_{\mu\nu\rho\sigma} \langle [ T^{\mu\alpha},f_{+}^{\rho\sigma} ] \nabla_\alpha u^\nu \rangle~, \nonumber \\
\mathcal{O}_{\rm TJP}^2 &=& i\varepsilon_{\mu\nu\rho\sigma} \langle [ \nabla^\nu T^\mu_\alpha,f_{+}^{\rho\sigma}] u^\alpha \rangle~, \nonumber \\
\mathcal{O}_{\rm TJP}^3 &=& i\varepsilon_{\mu\nu\rho\sigma} \langle [\nabla^\nu T^\mu_\alpha,f_{+}^{\rho\alpha}] u^\sigma \rangle~,    \label{Eq:TJP} \\[3mm]
\mathcal{O}_{\rm TVP}^1 &=& i\varepsilon_{\mu\nu\rho\sigma} \langle [ T^{\mu\alpha},V^{\rho\sigma} ] \nabla_\alpha u^\nu \rangle~, \nonumber \\
\mathcal{O}_{\rm TVP}^2 &=& i\varepsilon_{\mu\nu\rho\sigma} \langle [ \nabla^\nu T^\mu_\alpha,V^{\rho\sigma} ] u^\alpha \rangle~, \nonumber \\
\mathcal{O}_{\rm TVP}^3 &=& i\varepsilon_{\mu\nu\rho\sigma} \langle [ \nabla^\nu T^\mu_\alpha,V^{\rho\alpha} ] u^\sigma \rangle~.      \label{Eq:TVP}
\end{eqnarray}

\subsection{Observables for $e^+e^- \to K\bar{K}\pi$} \label{sec:3}
Following Ref.~\cite{Dai:2013joa}, the Feynman diagrams are built according to $e^+(q_1)e^-(q_2) \to \gamma^*(q) \to K(p_1)\bar{K}(p_2)\pi(p_3)$, and the amplitudes can be written as
\begin{eqnarray}
    \mathcal{M} = -\frac{4\pi\alpha}{q^2} i F_R(q^2,s,t) \varepsilon_{\mu\nu\alpha\beta} p_1^\nu p_2^\alpha p_3^\beta \bar{v}(q_1) \gamma^\mu u(q_2)~, \nonumber \\
    \label{Eq:M}
\end{eqnarray}
where $q=p_1+p_2+p_3$ is the c.m.f. energy and $F_R(q^2,s,t)$ is the relevant form factor. 
The Mandelstam variables are defined as $s=M_{K\bar{K}}^2=(p_1+p_2)^2$, $t=M_{\bar{K}\pi}^2=(p_2+p_3)^2$, and $u=M_{K\pi}^2=(p_1+p_3)^2$. Since $s+t+u=q^2+m_K^2+m_{\bar{K}}^2+m_\pi^2$, only three variables are independent. Notice that the masses of $m_K$, $m_{\bar{K}}$, and $m_\pi$ in the form factors are taken as the averaged values of the charged and neutral mesons. Their physical masses will be inputted into the phase space. See the next paragraph. 

Since we focus on the energy region from the $K\bar{K}\pi$ threshold, it is safe to ignore the mass of the electron. The cross-section for $e^+e^- \to K\bar{K}\pi$ can be expressed as
\begin{eqnarray} \label{eq:sigma}
    \sigma(q^2) = \frac{\alpha^2}{192\pi q^6} \int_{s_-}^{s_+}ds \int_{t_-}^{t_+}dt ~ \phi(q^2,s,t) \left| F_R(q^2,s,t) \right|^2~, \nonumber \\
\end{eqnarray}
where $\phi(q^2,s,t)$ is the three body phase space function of $K\bar{K}\pi$,
\begin{eqnarray}
    \phi(q^2,s,t) &=& -m_{\bar{K}}^2(m_{\bar{K}}^2 \!+\! q^2 \!-\! s \!-\! t)^2 \!+\! (-m_K^2 \!-\! m_{\bar{K}}^2 \!+\! s) \nonumber \\
    && \times (m_{\bar{K}}^2 \!+\! m_\pi^2 \!-\! t)(-m_{\bar{K}}^2 \!-\! q^2 \!+\! s \!+\! t) \nonumber \\
    && -m_\pi^2(m_K^2 \!+\! m_{\bar{K}}^2 \!-\! s)^2 \!-\! m_K^2(m_{\bar{K}}^2 \!+\! m_\pi^2 \!-\! t)^2 \nonumber \\
    && +4 m_K^2 m_{\bar{K}}^2 m_\pi^2~, \nonumber
\end{eqnarray}
and $s_\pm$/$t_\pm$ are the integration limits,
\begin{eqnarray}
    s_- &=& (m_K+m_{\bar{K}})^2~, ~ ~ ~ s_+ = \left( \sqrt{q^2}-m_\pi \right)^2~, \nonumber \\
    t_\pm &=& \frac{1}{4s} \bigg\{ (q^2-m_K^2+m_{\bar{K}}^2-m_\pi^2)^2 - [ \lambda^{1/2}(q^2,s,m_\pi^2) \nonumber \\
    && \mp \lambda^{1/2}(s,m_K^2,m_{\bar{K}}^2) ]^2 \bigg\}~, \nonumber
\end{eqnarray}
with $\lambda(a,b,c)=(a+b-c)^2-4ab$. 
The masses of the kaon and pion applied in the phase space function $\phi(q^2,s,t)$ and the upper and lower limits, $s_\pm$ and $t_\pm$, are the physical ones. 
The invariant mass spectra of $M_{K\bar{K}}=\sqrt{s}$ and $M_{\bar{K}\pi}=\sqrt{t}$ can be obtained from Eq.~(\ref{eq:sigma}), that is
\begin{eqnarray} \label{eq:dsigma}
    \frac{d\sigma}{d\sqrt{s}}(q^2,\sqrt{s}) \! &=& \! \frac{\alpha^2\sqrt{s}}{96\pi q^6} \int_{t_-}^{t_+}dt~\phi(q^2,s,t) \left| F_R(q^2,s,t) \right|^2\,, \nonumber \\
    \frac{d\sigma}{d\sqrt{t}}(q^2,\sqrt{t}) \! &=& \! \frac{\alpha^2\sqrt{t}}{96\pi q^6} \int_{s_-}^{s_+}ds~\phi(q^2,s,t) \left| F_R(q^2,s,t) \right|^2\,. \nonumber \\
\end{eqnarray}
In addition, normalization constants ($N$) should be multiplied to the differential cross section to compensate for the unknown efficiencies of the experimental events distribution data sets.

\subsection{Form factors of $e^+e^- \to K\bar{K}\pi$}
\begin{figure}[htb]
    \centering
    \includegraphics[width=0.50\textwidth]{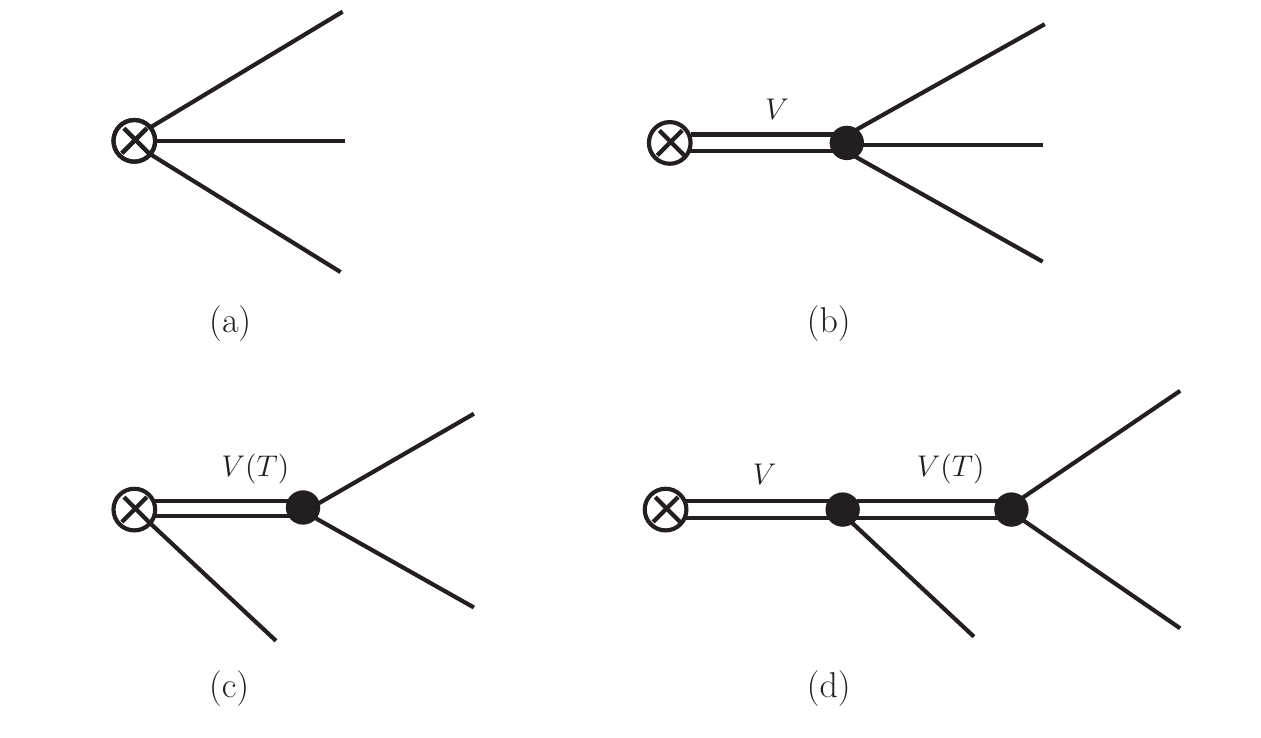}
    \caption{Diagrams contributing to the processes of $e^+e^- \to K\bar{K}\pi$ at leading order in the $1/N_C$ expansion. (a) The Wess-Zumino-Witten anomaly term coming from $\mathcal{L}_{(4)}^{\rm GB}$; (b), (c), and (d) are related to the contributions of vector and tensor resonances.}
    \label{fig:1}
\end{figure}

The Feynman diagrams contributing to $e^+e^- \to K\bar{K}\pi$ at leading order in the $1/N_C$ expansion is shown in Fig.~\ref{fig:1}.
The chiral anomaly term $\mathcal{L}^{\rm WZW}_{(4)}$ drives Fig.~\ref{fig:1} (a), and the electromagnetic current (virtual photon) will couple to the pseudoscalars directly. Fig.~\ref{fig:1} (b) is from the $\mathcal{O}_{\rm VPPP}^{j}$. Fig.~\ref{fig:1} (c) is related to the $\mathcal{O}_{\rm VJP}^{j}$, $\mathcal{L}^{\rm V}_{(2)}$, $\mathcal{O}_{\rm TJP}^{j}$, and $\mathcal{L}^{\rm T}_{(2)}$ terms, and Fig.~\ref{fig:1} (d) is related to the $\mathcal{O}_{\rm VVP}^{j}$, $\mathcal{L}^{\rm V}_{(2)}$, $\mathcal{O}_{\rm TVP}^{j}$, and $\mathcal{L}^{\rm T}_{(2)}$ terms. 
Compared with those diagrams for the processes of $e^+e^- \to \pi\pi\pi$, $\pi\pi\eta$~\cite{Dai:2013joa, Qin:2020udp}, the difference is that now we have extra vertexes involving tensors, i.e., the VJP, VVP and VPP vertexes can be replaced by TJP, TVP and TPP vertexes, respectively, See Figs.~\ref{fig:1} (c) and (d). 
The form factors are given as
\begin{equation} \label{eq:fr}
    F^j_R(q^2,s,t) = F^j_a + F^j_b + F^j_c + F^j_d~.
\end{equation}
Here, the subscripts \lq $a, b, c, d$' correspsond to the Feynman diagrams, and the superscript \lq $j=1, 2, 3$' denotes for $K^+K^-\pi^0$, $K^0_SK^0_L\pi^0$, and $K^0_SK^\pm\pi^\mp$ channels, respectively. 
The form factors for the process of $e^+e^- \to K^+(p_1)K^-(p_2)\pi^0(p_3)$ are given as
\begin{eqnarray} \label{eq:kpkmpi0}
    F_a^1 &=& -\frac{N_C}{12\pi^2F^3}~, \nonumber \\
    F_b^1 &=& \frac{2\sqrt{2}F_V}{F^3M_V} \bigg\{ \big[ \frac{1}{M_\rho^2 \!-\! q^2} \!+\! \frac{1}{3(M_\omega^2 \!-\! q^2)} \big] GR_1(q^2,s) \nonumber \\
    && -\frac{2 GR_2(q^2,s)}{3(M_\phi^2 \!-\! q^2)} \bigg\}~, \nonumber \\
    F_c^1 &=& -\frac{2\sqrt{2}G_V}{3F^3M_V} \bigg\{ \big[ \frac{1}{M_\rho^2 \!-\! s} \!+\! \frac{3}{M_\omega^2 \!-\! s} \big] CR_1(q^2,s,m_\pi^2) \nonumber \\
    && +\big[ \frac{CR_2(q^2,t)}{M_{K^\ast}^2 \!-\! t} \!+\! t \leftrightarrow u \big] \bigg\} \nonumber \\
    && +\frac{2g_T}{F^3M_{K_2^\ast}^2} \bigg\{ \frac{CR^T(q^2,s,t)}{M_{K_2^\ast}^2 \!-\! t} \!+\! t \leftrightarrow u \bigg\}~, \nonumber \\
    F_d^1 &=& \frac{2F_VG_V}{F^3} \bigg\{ \big[ \frac{2 DR(q^2,s,m_\pi^2)}{(M_\rho^2 \!-\! q^2)(M_\omega^2 \!-\! s)} \!+\! \frac{1}{3} \big( M_\rho^2 \leftrightarrow M_\omega^2 \big) \big] \nonumber \\
    && +\big[ \frac{1}{M_\rho^2 \!-\! q^2} \!+\! \frac{1}{3(M_\omega^2 \!-\! q^2)} \!-\! \frac{2}{3(M_\phi^2 \!-\! q^2)} \big] \nonumber \\
    && \times \big[ \frac{DR(q^2,m_K^2,t)}{M_{K^\ast}^2 \!-\! t} \!+\! t \leftrightarrow u \big] \bigg\} \nonumber \\
    && -\frac{F_Vg_T}{\sqrt{2}F^3M_{K_2^\ast}^2} \bigg\{ \big[ \frac{1}{M_\rho^2 \!-\! q^2} \!+\! \frac{1}{3(M_\omega^2 \!-\! q^2)} \nonumber \\
    && +\frac{2}{3(M_\phi^2 \!-\! q^2)} \big] \big[ \frac{DR^T(q^2,s,t)}{M_{K_2^\ast}^2 \!-\! t} \!+\! t \leftrightarrow u \big] \bigg\}~,
\end{eqnarray}
where the functions $GR_{1,2}$, $CR_{1,2}$, $DR$, $CR^T$ and $DR^T$ are defined in the Appendix~\ref{App:3}. 
Similarly, the form factors for $e^+e^- \to K^0_S(p_1)K^0_L(p_2)\pi^0(p_3)$ process are derived as
\begin{eqnarray} \label{eq:ksklpi0}
    F_a^2 &=& \frac{N_C}{12\pi^2F^3}~, \nonumber \\
    F_b^2 &=& -\frac{2\sqrt{2}F_V}{F^3M_V} \bigg\{ \big[ \frac{1}{M_\rho^2 \!-\! q^2} \!-\! \frac{1}{3(M_\omega^2 \!-\! q^2)} \big] GR_1(q^2,s) \nonumber \\
    && +\frac{2 GR_2(q^2,s)}{3(M_\phi^2 \!-\! q^2)} \bigg\}~, \nonumber \\
    F_c^2 &=& -\frac{2\sqrt{2}G_V}{3F^3M_V} \bigg\{ \big[ \frac{1}{M_\rho^2 \!-\! s} \!-\! \frac{3}{M_\omega^2 \!-\! s} \big] CR_1(q^2,s,m_\pi^2) \nonumber \\
    && -2\big[ \frac{CR_1(q^2,t,m_K^2)}{M^2_{K^\ast} \!-\! t} \!+\! t \leftrightarrow u \big] \bigg\}~, \nonumber \\
    F_d^2 &=& -\frac{2F_VG_V}{F^3} \bigg\{ \big[ \frac{2 DR(q^2,s,m_\pi^2)}{(M_\rho^2 \!-\! q^2)(M_\omega^2 \!-\! s)} \!-\! \frac{1}{3} \big( M_\rho^2 \leftrightarrow M_\omega^2 \big) \big] \nonumber \\
    && +\big[ \frac{1}{M_\rho^2 \!-\! q^2} \!-\! \frac{1}{3(M_\omega^2 \!-\! q^2)} \!+\! \frac{2}{3(M_\phi^2 \!-\! q^2)} \big] \nonumber \\
    && \times \big[ \frac{DR(q^2,t,m_K^2)}{M_{K^\ast}^2 \!-\! t} \!+\! t \leftrightarrow u \big] \bigg\} \nonumber \\
    && +\frac{F_Vg_T}{\sqrt{2}F^3M_{K_2^\ast}^2} \bigg\{ \big[ \frac{1}{M_\rho^2 \!-\! q^2} \!-\! \frac{1}{3(M_\omega^2 \!-\! q^2)} \nonumber \\
    && -\frac{2}{3(M_\phi^2 \!-\! q^2)} \big] \big[ \frac{DR^T(q^2,s,t)}{M_{K_2^\ast}^2 \!-\! t} \!+\! t \leftrightarrow u \big] \bigg\}~.
\end{eqnarray}
Notice that the tensor part of Fig.~\ref{fig:1} (c) does not have contributions to this process. 
The form factors of $K^0_SK^+\pi^-$ and $K^0_SK^-\pi^+$ differ by one overall phase, so we only need to know one of the form factors in these two processes, e.g., $e^+e^- \to K^0_S(p_1)K^+(p_2)\pi^-(p_3)$. One has
\begin{eqnarray} \label{eq:kskpi}
    F_a^3 &=& 0~, \nonumber \\
    F_b^3 &=& \frac{2\sqrt{2}F_V}{F^3M_V} \bigg\{ \frac{(g_1 \!+\! 2g_2 \!-\! g_3)(t \!-\! u)}{M_\rho^2 \!-\! q^2} \nonumber \\
    && -\frac{GR_1(q^2,s)}{3(M_\omega^2 \!-\! q^2)} \!+\! \frac{2 GR_2(q^2,s)}{3(M_\phi^2 \!-\! q^2)} \bigg\}~, \nonumber \\
    F_c^3 &=& \frac{2\sqrt{2}G_V}{3F^3M_V} \bigg\{ \frac{CR_1(q^2,s,m_\pi^2)}{M_\rho^2 \!-\! s} \!-\! \frac{2 CR_1(q^2,t,m_K^2)}{M^2_{K^\ast} \!-\! t} \nonumber \\
    && +\frac{CR_2(q^2,u)}{M^2_{K^\ast} \!-\! u} \bigg\} \!-\! \frac{2g_T}{F^3} \bigg\{ \frac{(2c_1^T \!-\! 2c_2^T \!-\! c_3^T)(t \!-\! u)}{M_{a_2}^2 \!-\! s} \nonumber \\
    && +\frac{CR^T(q^2,s,u)}{M_{K_2^\ast}^2(M_{K_2^\ast}^2 \!-\! u)} \bigg\}~, \nonumber \\
    F_d^3 &=& -\frac{2F_VG_V}{F^3} \bigg\{ \frac{2 DR(q^2,s,m_\pi^2)}{3(M_\omega^2 \!-\! q^2)(M_\rho^2 \!-\! s)} \nonumber \\
    && -\big[ \frac{1}{M_\rho^2 \!-\! q^2} \!-\! \frac{1}{3(M_\omega^2 \!-\! q^2)} \!+\! \frac{2}{3(M_\phi^2 \!-\! q^2)} \big] \nonumber \\
    && \times \frac{DR(q^2,t,m_K^2)}{M_{K^\ast}^2 \!-\! t} \nonumber \\
    && +\big[ \frac{1}{M_\rho^2 \!-\! q^2} \!+\! \frac{1}{3(M_\omega^2 \!-\! q^2)} \!-\! \frac{2}{3(M_\phi^2 \!-\! q^2)} \big] \nonumber \\
    && \times \frac{DR(q^2,u,m_K^2)}{M_{K^\ast}^2 \!-\! u} \bigg\} \nonumber \\
    && +\frac{F_Vg_T}{\sqrt{2}F^3} \bigg\{ \frac{2(2d_1^T \!-\! 2d_2^T \!-\! d_3^T)(t \!-\! u)}{(M_\rho^2 \!-\! q^2)(M_{a_2}^2 \!-\! s)} \nonumber \\
    && -\frac{1}{M_{K_2^\ast}^2} \big[ \big( \frac{1}{M_\rho^2 \!-\! q^2} \!-\! \frac{1}{3(M_\omega^2 \!-\! q^2)} \!-\! \frac{2}{3(M_\phi^2 \!-\! q^2)} \big) \nonumber \\
    && \times \frac{DR^T(q^2,s,t)}{M_{K_2^\ast}^2 \!-\! t} \nonumber \\
    && -\big( \frac{1}{M_\rho^2 \!-\! q^2} \!+\! \frac{1}{3(M_\omega^2 \!-\! q^2)} \!+\! \frac{2}{3(M_\phi^2 \!-\! q^2)} \big) \nonumber \\
    && \times \frac{DR^T(q^2,s,u)}{M_{K_2^\ast}^2 \!-\! u} \big] \bigg\}~.
\end{eqnarray}
Notice that for the processes with final states $K^0_S(p_1)K^\pm(p_2)\pi^\mp(p_3)$, Fig.~\ref{fig:1} (a) has no contribution, i.e., the Wess-Zumino-Witten anomaly term does not contribute. One needs to add the two cross sections, $\sigma(K^0_SK^\pm\pi^\mp)=\sigma(K^0_SK^+\pi^-)+\sigma(K^0_SK^-\pi^+)=2\sigma(K^0_SK^+\pi^-)$ to calculate the HVP contributions to the $(g-2)_\mu$.

\subsection{Constraints on the form factors}
There are dozens of unknown couplings coming from the effective Lagrangians that need to be fixed, as shown in the form factors of Eqs.~(\ref{eq:kpkmpi0}, \ref{eq:ksklpi0}, \ref{eq:kskpi}). One can match the Green functions between RChT and QCD in the high energy region to solve this problem~\cite{Ruiz-Femenia:2003jdx, Dai:2019lmj}. 
Here, we use the constraints obtained by Ref.~\cite{Dai:2013joa} that focus on the processes with final states $\pi\pi\pi$ and $\pi\pi\eta$. In that analysis, demanding the two-point Green function of the vector current (with the contributions from the exclusive channels of $\pi\pi\pi$ and $\pi\pi\eta$) to vanish in the energy region $q^2 \to \infty$ gives
\begin{eqnarray} \label{Eq:constraint;3pi}
    g_1+2g_2-g_3 &=& 0~, \nonumber \\
    g_2 &=& \frac{N_C M_V}{192\sqrt{2} \pi^2 F_V}~, \nonumber \\
    c_1-c_2+c_5 &=& 0~, \nonumber \\
    c_1+c_2+8c_3-c_5 &=& 0~, \nonumber \\
    c_1-c_2-c_5+2c_6 &=& -\frac{N_C M_V}{96\sqrt{2} \pi^2 G_V}~, \nonumber \\
    d_3 &=& -\frac{N_C M_V^2}{192 \pi^2 F_V G_V}~.
\end{eqnarray}
Moreover, if we adopt the same scheme as Ref.~\cite{Dai:2013joa} to get short distance constraints in the process of $e^+e^- \to K\bar{K}\pi$, the matching procedure gives two extra constraints for the tensor couplings,
\begin{eqnarray}
    2c_1^T - 2c_2^T - c_3^T &=& 0~, \nonumber \\
    2d_1^T - 2d_2^T - d_3^T &=& 0~. \nonumber
\end{eqnarray}
Unfortunately, the widths of $\Gamma(a_2^\pm \to \pi^\pm \gamma)$, $\Gamma(K_2^\ast \to K^\ast \pi)$, $\Gamma(K_2^\ast \to \rho(\omega) K)$, and $\Gamma(K_2^{\ast\pm} \to K^\pm \gamma)$ would vanish if we apply these short distance constraints, which is incompatible with the experimental data~\cite{ParticleDataGroup:2022pth}. See discussions in the next sections. 
This conflict may be due to the lack of insight into the matching between RChT and QCD involving tensor current. Therefore, we leave $2c_1^T-2c_2^T-c_3^T$ and $2d_1^T-2d_2^T-d_3^T$ free, but determine them with the fit to the experiment data for the tensor resonances. In addition, $g_4$ and $c_4$ have been studied in the analysis of $\tau \to KK\pi\nu_\tau$~\cite{Dumm:2009kj}, which could be a guide to the present analysis. However, further study is still needed on the necessary isospin-breaking corrections, which limits the reference value of the couplings determined from $\tau$ decay. 
We also take the short distance constraints obtained in other analyses, such as those from the two-pion vector form factor and from the study of matching the three-point $\langle VVP \rangle$ Green functions between QCD and RChT~\cite{Ruiz-Femenia:2003jdx},
\begin{eqnarray} \label{Eq:constraint;2pi}
    F_V G_V = F^2~, ~ ~ ~ d_1+8d_2-d_3 = \frac{F^2}{8F_V^2}~.
\end{eqnarray}
With these constraints, we have reduced the unknown coupling constants, and only a few of them are left, i.e., $F_V$, $g_4$, $c_4$, $2g_4+g_5$, $g_T$, $c_3^T$, $d_3^T$, $2c_1^T-2c_2^T-c_3^T$, and $2d_1^T-2d_2^T-d_3^T$. 

The interaction Lagrangians discussed above are only for the lightest vector and tensor mesons, which dominate the interactions around 1 GeV. However, the heavier states appear in the higher energy region and would contribute, too. 
In order to account for the excited resonance effects up to the energy region that we studied, $E_{\rm cm} \sim 2.3$ GeV, we adopt the same strategy as done in Refs.~\cite{Dai:2013joa, Qin:2020udp, Wang:2023njt} to deal with the heavier states. 
Here, we include two multiplets of the vector resonances ($V'_{\mu\nu}$ and $V_{\mu\nu}''$) and one multiplet of tensor resonances ($T'_{\mu\nu}$), since the second excited tensor resonances lie above the energy region we focus on. These heavier multiplets are included by extension of the Breit-Wigner propagator~\cite{Dai:2013joa}
\begin{eqnarray} \label{eq:bw}
    \frac{1}{M_R^2-x} \to \frac{1}{M_R^2-x} + \frac{\beta^{R'}_j}{M_{R'}^2-x} + \frac{\beta^{R''}_j}{M_{R''}^2-x}~,
\end{eqnarray}
where $R=V,T$ and $\beta^{T''}_j=0$. The subscript \lq $j=1, 2, 3$' represents the $K^+K^-\pi^0$, $K^0_SK^0_L\pi^0$, and $K^0_SK^\pm\pi^\mp$ channels, respectively. 
Indeed, assuming that one writes down these heavier fields explicitly in the chiral effective Lagrangians, the Feynman diagrams will have the identical topologies as those given in Fig.~\ref{fig:1}, and the form factors will have similar formalism with only some energy functions absorbed into the parameters of $\beta_{j}^{R',R''}$. 
For convenience, we collect the lightest vector and tensor resonances $R$ and their heavier partners $R'$, $R''$ that are used in the present analysis,
\begin{eqnarray}
    R \!&=&\! \big\{ \rho(770), \omega(782), \phi(1020), K^\ast(892), \nonumber \\
    && ~ ~ a_2(1320), K_2^\ast(1430) \big\}~, \nonumber \\
    R' \!&=&\! \big\{ \rho(1450), \omega(1420), \phi(1680), K^\ast(1410), \nonumber \\
    && ~ ~ a_2(1700), K_2^\ast(1980) \big\}~, \nonumber \\
    R'' \!&=&\! \big\{ \rho(1700), \omega(1650), \phi(2170), K^\ast(1680) \big\}~. \nonumber
\end{eqnarray}
In addition, one would notice that the propagator in the form factors of Eq.~(\ref{eq:fr}) are real, while in the real world, the resonances have unignorable widths. Also, the energy-dependent widths can give a better description of the data. 
These are fulfilled by applying the well-known Breit-Wigner propagators
\begin{eqnarray} \label{eq:gamma}
    \frac{1}{M_R^2-x} \to \frac{1}{M_R^2-x-iM_R\Gamma_R(x)}~,
\end{eqnarray}
where the widths of $\rho(770)$ and $K^\ast(892)$ and their corresponding excited states are taken in the energy-dependent form. See Appendix~\ref{App:4}. For the other resonances, constant widths from PDG~\cite{ParticleDataGroup:2022pth} are adopted.

Besides, in the high energy region, the form factors will be divergent, as the terms of $m_K^2$ (multiplied together with the momentum) have significant contributions, while it is ignored in the chiral limit when the high energy constraints are obtained. 
To fix this problem, we follow the method applied in Ref.~\cite{Wang:2023njt} and implement a regulator $f(q^2)={\rm exp} \left\{ -q^6/\Lambda^6 \right\}$~\cite{Yang:2022qoy} in the form factors, where the cut-off is chosen as $\Lambda=2.3, 2.4, 2.5, 2.6, 2.7$ GeV. As will be discussed in the following sections, we choose $2.5$ GeV as the optimal one.

In order to take more constraints from the experiment, the correlated decay processes of the vector meson, $K^\ast$, and tensor mesons, $a_2$ and $K_2^\ast$, are also calculated, that is, $K^\ast \to PP/P\gamma$, $a_2 \to PP/P\gamma$ and $K_2^\ast \to PP/VP/P\gamma$, where $P$ is the pseudoscalar. Here, one has $K^\ast=K^\ast(892)$, $a_2=a_2(1320)$ and $K_2^\ast=K_2^\ast(1430)$. The complete expressions for these decay widths are collected in Appendix~\ref{App:5}. 
What is more, the mixing angle $\theta_P$ will affect the decay widths of $a_2 \to \eta\pi$, $\eta'\pi$, and it should also be determined by the fit. Also, there need three normalization factors, $N_1, N_2, N_3$, which are multiplied by the differential cross section to fit the events distribution of the invariant mass spectra. 
Finally, the processes of $e^+e^- \to K\bar{K}\pi$ and the correlated decays of the resonances are studied within RChT. After taking the short distance constraints, there are still some parameters left to be fixed by the experimental data, i.e., $F_V$, $g_4$, $c_4$, $2g_4+g_5$, $g_T$, $c_3^T$, $d_3^T$, $2c_1^T-2c_2^T-c_3^T$, $2d_1^T-2d_2^T-d_3^T$, $\beta^{V'}_{j}$, $\beta^{V''}_{j}$, $\beta^{T'}_{j}$, and the mixing angle $\theta_P$ of $\eta-\eta'$. In addition, the masses and widths of the heavier resonances are restricted by PDG~\cite{ParticleDataGroup:2022pth}.

\section{Numerical results} \label{sec:4}
To reach a comprehensive analysis and obtain reliable form factors, we fit our amplitudes to all the datasets of the cross sections, two-body invariant mass spectra, and decay widths of vectors and tensors. The cross sections of the processes of electron-positron annihilation into $K\bar{K}\pi$ ($K^+K^-\pi^0$, $K^0_SK^0_L\pi^0$ and $K^0_SK^\pm\pi^\mp$) are measured by Refs.~\cite{Mane:1982si, Bisello:1991kd, BaBar:2007ceh, Solodov:2016klh, BaBar:2017nrz, Achasov:2017vaq, Semenov:2019zzi, Uskov:2020xna, SND:2020qmb, BESIII:2022wxz, BESIII:2023xac}. 
We performed an overall fit using all the datasets to determine the unknown parameters. As is well known, the angular distribution data sets are helpful for constraining the amplitude~\cite{Dai:2014zta, Dai:2014lza}. Here, there are three kinds of datasets for the invariant mass spectra~\cite{SND:2020qmb, Achasov:2017vaq}, ($M_{K^\pm\pi^0}$, $M_{K_{S/L}^0\pi^0}$ and $M_{K_S^0K_L^0}$) for the relevant processes of $e^+e^- \to K\bar{K}\pi$, which are included in our fit to refine the present analysis, as $t,u$ are functions of the Mandelstam variable $s$ and the scattering angle $\theta_s$. 
Most of these invariant mass spectra are in the energy region of [$1.5-1.75$] GeV, which can help check the reliability of our model, where the generalizing propagators are applied to include the contributions of the heavier resonances. 
We summarize all the datasets in Table~\ref{tab:data}.
\begin{table}[htb]
    \centering
    \renewcommand\arraystretch{1.2}
    \setlength{\tabcolsep}{8pt}
    \scalebox{1}{
    \begin{tabular}{c c c c}
    \hline
    \hline
    Collaboration  &  $K^+K^-\pi^0$  &  $K^0_SK^0_L\pi^0$  &  $K^0_SK^\pm\pi^\mp$ \\
    \hline
    DM1     &  $-$                     &  $-$                     &  \cite{Mane:1982si}    \\
    DM2     &  \cite{Bisello:1991kd}   &  $-$                     &  \cite{Bisello:1991kd} \\
    BABAR   &  \cite{BaBar:2007ceh}    &  \cite{BaBar:2017nrz}    &  \cite{BaBar:2007ceh}  \\
    SND     &  \cite{SND:2020qmb}      &  \cite{Achasov:2017vaq}  &  $-$ \\
    CMD-3   &  \cite{Solodov:2016klh}  &  \cite{Semenov:2019zzi}  &  \cite{Uskov:2020xna}  \\
    BESIII  &  \cite{BESIII:2022wxz}   &  \cite{BESIII:2023xac}   &  $-$ \\
    \hline
    \hline
    \end{tabular}}
    \caption{The datasets of the process of $e^+e^- \to K\bar{K}\pi$ adopted for the present analysis.}
    \label{tab:data}
\end{table}

We list the fit results with different cut-offs in Table~\ref{tab:para;cut}.
\begin{table*}[htb]
    \centering
    \renewcommand\arraystretch{1.3}
    \setlength{\tabcolsep}{10pt}
    \scalebox{0.84}{
    \begin{tabular}{c  @{}  c  c  c  c  c  c  @{}}
    \hline
    \hline
    Parameter                        &    $\Lambda=2.3$ GeV     &    $\Lambda=2.4$ GeV     &    $\Lambda=2.5$ GeV     &    $\Lambda=2.6$ GeV     &    $\Lambda=2.7$ GeV    &    PDG~\cite{ParticleDataGroup:2022pth}  \\
    \hline
    $F_V$ (GeV)                      &    $ 0.1480\pm0.0001$    &    $ 0.1480\pm0.0001$    &    $ 0.1480\pm0.0001$    &    $ 0.1480\pm0.0001$    &    $ 0.1480\pm0.0001$   &                     \\
    $g_4$                            &    $-0.0233\pm0.0017$    &    $-0.0260\pm0.0024$    &    $-0.0470\pm0.0024$    &    $ 0.0005\pm0.0021$    &    $ 0.0017\pm0.0027$   &                     \\
    $c_4$                            &    $-0.0003\pm0.0001$    &    $-0.0004\pm0.0001$    &    $-0.0004\pm0.0001$    &    $-0.0004\pm0.0001$    &    $-0.0004\pm0.0001$   &                     \\
    $2g_4+g_5$                       &    $-0.0235\pm0.0020$    &    $-0.0326\pm0.0047$    &    $-0.0294\pm0.0047$    &    $-0.0553\pm0.0029$    &    $-0.0625\pm0.0032$   &                     \\
    $g_T$ (GeV)                      &    $ 0.0235\pm0.0001$    &    $ 0.0233\pm0.0002$    &    $ 0.0233\pm0.0001$    &    $ 0.0233\pm0.0002$    &    $ 0.0233\pm0.0001$   &                     \\
    $c_3^T$ (GeV)                    &    $ 0.0800\pm0.0294$    &    $ 0.1827\pm0.0305$    &    $ 0.3647\pm0.0271$    &    $ 0.2538\pm0.0293$    &    $ 0.2216\pm0.0255$   &                     \\
    $d_3^T$                          &    $-0.5729\pm0.2385$    &    $-0.7223\pm0.1297$    &    $-0.7862\pm0.0602$    &    $-0.7052\pm0.2707$    &    $-0.7835\pm1.0472$   &                     \\
    $2c_1^T-2c_2^T-c_3^T$ (GeV)      &    $ 0.0505\pm0.0014$    &    $ 0.0407\pm0.0010$    &    $ 0.0342\pm0.0008$    &    $ 0.0320\pm0.0008$    &    $ 0.0267\pm0.0007$   &                     \\
    $2d_1^T-2d_2^T-d_3^T$            &    $ 0.7022\pm0.0026$    &    $ 0.7022\pm0.0014$    &    $ 0.7022\pm0.0007$    &    $ 0.7022\pm0.0005$    &    $ 0.7022\pm0.0004$   &                     \\
    $\theta_P(^\circ)$               &    $ -18.00\pm0.31  $    &    $ -18.00\pm0.32  $    &    $ -18.00\pm0.32  $    &    $ -18.00\pm0.32  $    &    $ -18.00\pm0.32  $   &                     \\
    $\beta^{V'}_{1}$                 &    $-0.3212\pm0.0048$    &    $-0.2961\pm0.0019$    &    $-0.2685\pm0.0023$    &    $-0.3683\pm0.0022$    &    $-0.3576\pm0.0023$   &                     \\
    $\beta^{V''}_{1}$                &    $-0.0147\pm0.0016$    &    $-0.0267\pm0.0018$    &    $-0.0289\pm0.0017$    &    $-0.0224\pm0.0013$    &    $-0.0221\pm0.0012$   &                     \\
    $\beta^{V'}_{2}$                 &    $-0.5192\pm0.0022$    &    $-0.4713\pm0.0026$    &    $-0.4428\pm0.0019$    &    $-0.4423\pm0.0031$    &    $-0.4253\pm0.0037$   &                     \\
    $\beta^{V''}_{2}$                &    $ 0.0214\pm0.0055$    &    $ 0.0152\pm0.0037$    &    $ 0.0170\pm0.0035$    &    $ 0.0219\pm0.0029$    &    $ 0.0325\pm0.0041$   &                     \\
    $\beta^{V'}_{3}$                 &    $-0.4941\pm0.0036$    &    $-0.4546\pm0.0028$    &    $-0.4271\pm0.0029$    &    $-0.4509\pm0.0035$    &    $-0.4354\pm0.0037$   &                     \\
    $\beta^{V''}_{3}$                &    $ 0.0623\pm0.0074$    &    $ 0.0450\pm0.0088$    &    $ 0.0377\pm0.0080$    &    $ 0.0469\pm0.0051$    &    $ 0.0535\pm0.0053$   &                     \\
    $M_{\rho'}$ (GeV)                &    $ 1.5297\pm0.0017$    &    $ 1.5323\pm0.0019$    &    $ 1.5485\pm0.0018$    &    $ 1.4962\pm0.0015$    &    $ 1.4943\pm0.0020$   &    $1.465\pm0.025$  \\
    $\Gamma_{\rho'}$ (GeV)           &    $ 0.2331\pm0.0036$    &    $ 0.2218\pm0.0006$    &    $ 0.2215\pm0.0002$    &    $ 0.2215\pm0.0003$    &    $ 0.2215\pm0.0003$   &    $0.400\pm0.060$  \\
    $M_{\omega'}$ (GeV)              &    $ 1.4237\pm0.0033$    &    $ 1.4200\pm0.0040$    &    $ 1.4223\pm0.0041$    &    $ 1.2400\pm0.0033$    &    $ 1.2400\pm0.0027$   &    $1.410\pm0.060$  \\
    $\Gamma_{\omega'}$ (GeV)         &    $ 0.1439\pm0.0037$    &    $ 0.1597\pm0.0063$    &    $ 0.1791\pm0.0057$    &    $ 0.4800\pm0.0234$    &    $ 0.4800\pm0.0067$   &    $0.290\pm0.190$  \\
    $M_{\phi'}$ (GeV)                &    $ 1.6517\pm0.0009$    &    $ 1.6515\pm0.0009$    &    $ 1.6515\pm0.0009$    &    $ 1.6349\pm0.0007$    &    $ 1.6336\pm0.0007$   &    $1.680\pm0.020$  \\
    $\Gamma_{\phi'}$ (GeV)           &    $ 0.1814\pm0.0016$    &    $ 0.1642\pm0.0011$    &    $ 0.1635\pm0.0013$    &    $ 0.1582\pm0.0013$    &    $ 0.1533\pm0.0016$   &    $0.150\pm0.050$  \\
    $M_{K^{\ast'}}$ (GeV)            &    $ 1.3705\pm0.0049$    &    $ 1.3705\pm0.0183$    &    $ 1.4789\pm0.0738$    &    $ 1.4800\pm0.0746$    &    $ 1.4800\pm0.0134$   &    $1.414\pm0.015$  \\
    $\Gamma_{K^{\ast'}}$ (GeV)       &    $ 0.2104\pm0.0048$    &    $ 0.2312\pm0.0256$    &    $ 0.3500\pm0.0866$    &    $ 0.3500\pm0.0221$    &    $ 0.3500\pm0.0114$   &    $0.232\pm0.021$  \\
    $M_{\rho''}$ (GeV)               &    $ 1.7939\pm0.0009$    &    $ 1.7937\pm0.0043$    &    $ 1.7939\pm0.0026$    &    $ 1.7940\pm0.0008$    &    $ 1.7940\pm0.0005$   &    $1.720\pm0.020$  \\
    $\Gamma_{\rho''}$ (GeV)          &    $ 0.1945\pm0.0153$    &    $ 0.3842\pm0.0243$    &    $ 0.4259\pm0.0042$    &    $ 0.1502\pm0.0039$    &    $ 0.1501\pm0.0114$   &    $0.250\pm0.100$  \\
    $M_{\omega''}$ (GeV)             &    $ 1.7536\pm0.0232$    &    $ 1.7536\pm0.0901$    &    $ 1.7536\pm0.0289$    &    $ 1.7536\pm0.0914$    &    $ 1.6110\pm0.0845$   &    $1.670\pm0.030$  \\
    $\Gamma_{\omega''}$ (GeV)        &    $ 0.4240\pm0.0903$    &    $ 0.4215\pm0.0156$    &    $ 0.4236\pm0.0152$    &    $ 0.4232\pm0.0279$    &    $ 0.4251\pm0.0899$   &    $0.315\pm0.035$  \\
    $M_{\phi''}$ (GeV)               &    $ 2.1934\pm0.0092$    &    $ 2.1778\pm0.0051$    &    $ 2.1758\pm0.0047$    &    $ 2.1765\pm0.0056$    &    $ 2.1679\pm0.0052$   &    $2.162\pm0.007$  \\
    $\Gamma_{\phi''}$ (GeV)          &    $ 0.0877\pm0.0077$    &    $ 0.1033\pm0.0066$    &    $ 0.1119\pm0.0073$    &    $ 0.1302\pm0.0082$    &    $ 0.1605\pm0.0121$   &    $0.100^{+0.031}_{-0.023}$  \\
    $M_{K^{\ast''}}$ (GeV)           &    $ 1.6995\pm0.0705$    &    $ 1.7802\pm0.1094$    &    $ 1.7802\pm0.1020$    &    $ 1.7802\pm0.1059$    &    $ 1.7802\pm0.1098$   &    $1.718\pm0.018$  \\
    $\Gamma_{K^{\ast''}}$ (GeV)      &    $ 0.2800\pm0.1981$    &    $ 0.4804\pm0.2000$    &    $ 0.4805\pm0.1847$    &    $ 0.4805\pm0.1638$    &    $ 0.4805\pm0.1946$   &    $0.322\pm0.110$  \\
    $N_1$                            &    $24.7194\pm0.5964$    &    $25.4024\pm0.6281$    &    $26.6570\pm0.6002$    &    $25.5414\pm0.5542$    &    $26.1772\pm0.6214$   &                     \\
    $N_2$                            &    $ 3.4257\pm0.0913$    &    $ 3.3753\pm0.0929$    &    $ 3.3565\pm0.0947$    &    $ 3.3705\pm0.0895$    &    $ 3.3412\pm0.0937$   &                     \\
    $N_3$                            &    $ 2.1689\pm0.0703$    &    $ 2.1515\pm0.0705$    &    $ 2.1722\pm0.0711$    &    $ 2.1175\pm0.0670$    &    $ 2.1084\pm0.0689$   &                     \\
    \hline
    $\chi_{1}^{2} / {\rm d.o.f.}$    & $\frac{1135.64}{515-35}=2.37$ & $\frac{1193.52}{515-35}=2.49$ & $\frac{1304.51}{515-35}=2.72$ & $\frac{1378.13}{515-35}=2.87$ & $\frac{1496.68}{515-35}=3.12$ &   \\
    $\chi_{2}^{2} / {\rm d.o.f.}$    & $\frac{1036.22}{503-34}=2.21$ & $\frac{1054.05}{503-34}=2.25$ & $\frac{1115.62}{503-34}=2.38$ & $\frac{1167.30}{503-34}=2.49$ & $\frac{1220.60}{503-34}=2.60$ &   \\
    \hline
    \hline
\end{tabular}}
    \caption{Fitted parameters with different cut-offs, $\Lambda=2.3$ GeV, $2.4$ GeV, $2.5$ GeV, $2.6$ GeV and $2.7$ GeV. The uncertainties of the parameters are taken from MINUIT~\cite{James:1975dr}. $\chi_{1}^{2} / {\rm d.o.f.}$ is for all the data points and $\chi_{2}^{2} / {\rm d.o.f.}$ is for the data points without decaying widths.}
    \label{tab:para;cut}
\end{table*}
The fit quality decreases as the cut-off increases. Besides, the $f(q^2)$ function works like a step function and will have little contribution in the energy region below $\sqrt{s}=1.8, 1.9, 2.0, 2.1, 2.2$ GeV for the cut-offs of $\Lambda=2.3, 2.4, 2.5, 2.6, 2.7$ GeV, respectively. Balancing the fit quality and less effects on the physics in the energy region below 2 GeV, we choose $\Lambda=2.5$ GeV as the optimal one.

In practice, it is found that one can set $\beta^{T'}_{j}=\beta^{V'}_{j}$, and the results are almost as good as what is obtained by setting them free. 
In the present analysis, the magnitude of the parameter $2g_4+g_5$ is much smaller than what is obtained from the analyses on $e^+e^- \to \pi\pi\pi$, $\pi\pi\eta$~\cite{Dai:2013joa, Qin:2020udp}. 
The reason is that in the $\pi\pi\pi$ and $\pi\pi\eta$ cases, the $2g_4+g_5$ is multiplied with the mass term $m_\pi^2$, while we have $(2g_4+g_5)m_K^2$ here. See Eq.~(\ref{Eq:CRDR}). One would need a smaller $2g_4+g_5$ multiplying $m_K^2$ to have similar contributions with that of $(2g_4+g_5)m_\pi^2$. Indeed, this explanation is made from a phenomenological view. In principle, the unknown couplings should be consistent in different processes. Nevertheless, given the fact that the final state interactions (FSI) is partly accounted for in our current approach, these couplings may compensate for the incomplete treatment of FSI. 
The parameters obtained in the current analysis are compatible with those of other works such as Refs.~\cite{Dai:2013joa, Qin:2020udp, Wang:2023njt} and PDG, considering the corresponding uncertainties. 
The masses and widths of the ground states of vectors and tensors that are used in this analysis are fixed by PDG~\cite{ParticleDataGroup:2022pth}. 

\begin{figure*}[htbp]
    \centering
    \includegraphics[width=0.48\textwidth]{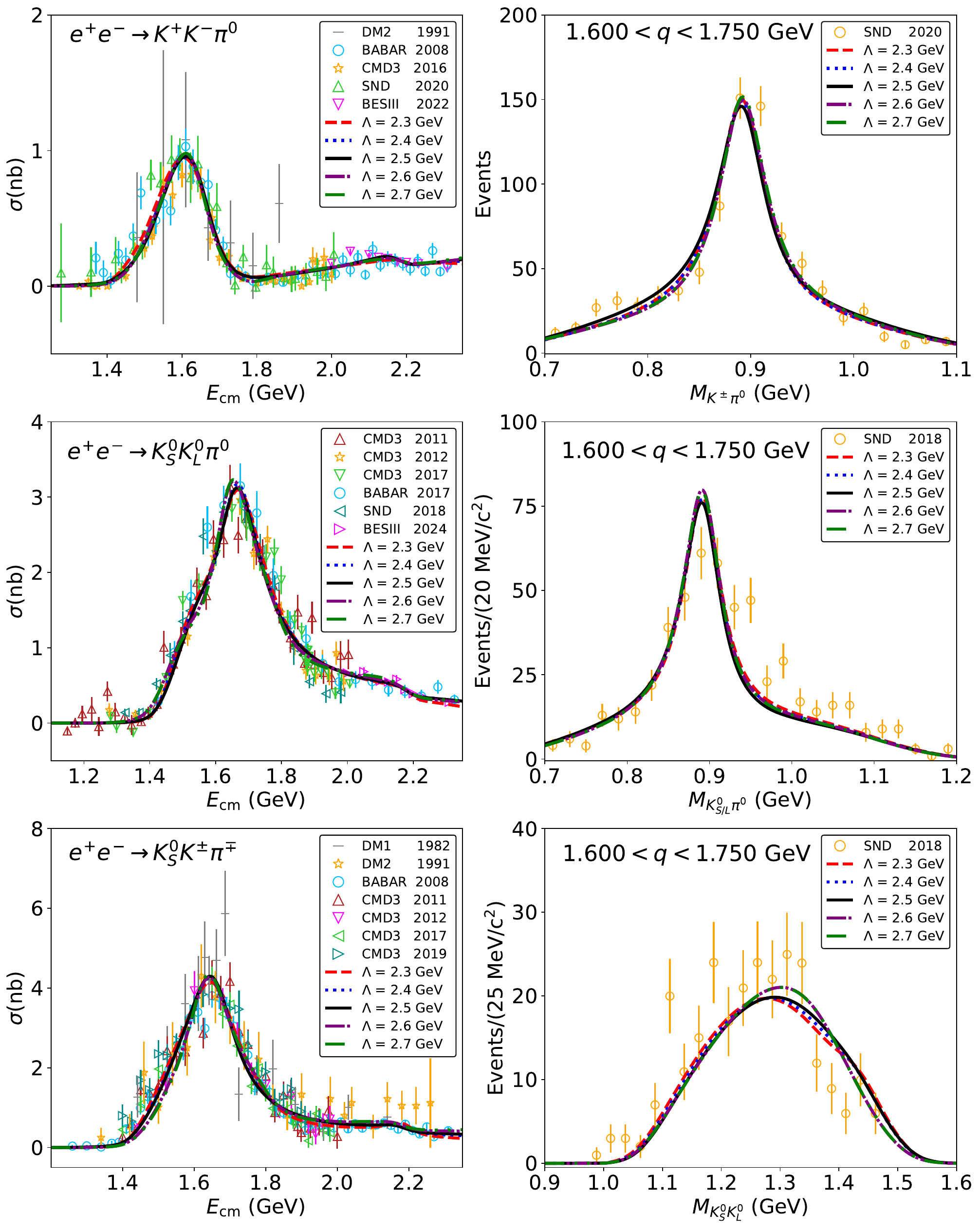} 
    \quad
    \includegraphics[width=0.48\textwidth]{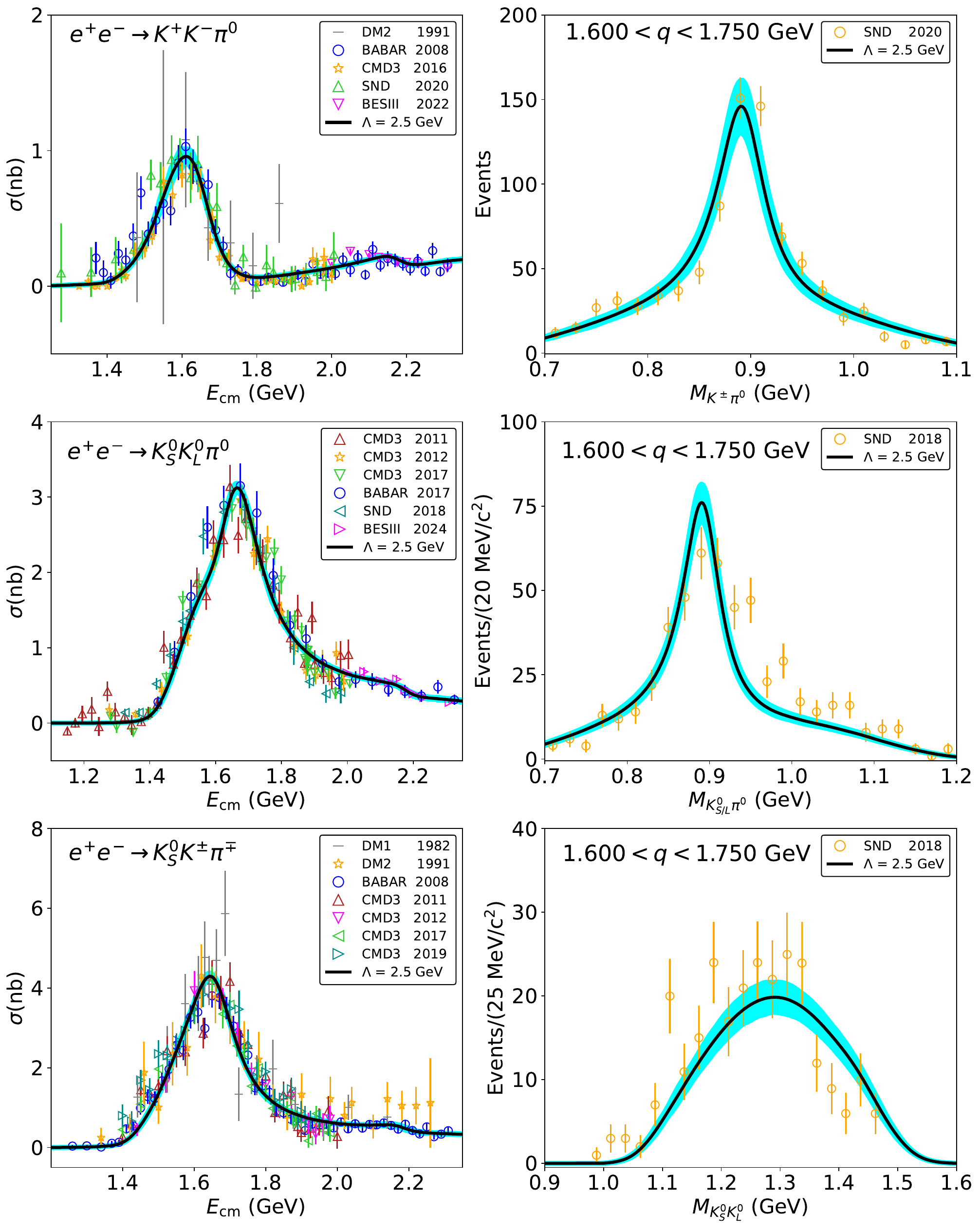}
    \caption{Fit to the cross sections and invariant mass spectra. The graphs of the two left columns are the results of different cut-offs, and the ones of the two right columns are for $\Lambda=2.5$ GeV. 
    The cross sections are of $e^+e^- \to K^+K^-\pi^0$, $K_S^0K_L^0\pi^0$ and $K_S^0K^\pm\pi^\mp$, where the data are from DM1~\cite{Mane:1982si}, DM2~\cite{Bisello:1991kd}, BABAR~\cite{BaBar:2007ceh, BaBar:2017nrz}, SND~\cite{Achasov:2017vaq, SND:2020qmb}, CMD-3~\cite{Solodov:2016klh, Semenov:2019zzi, Uskov:2020xna} and BESIII~\cite{BESIII:2022wxz, BESIII:2023xac}. The data of invariant mass spectra of $M_{K^\pm\pi^0}$, $M_{K_{S/L}^0\pi^0}$ and $M_{K_S^0K_L^0}$ from the $K^+K^-\pi^0$ and $K_S^0K_L^0\pi^0$ processes are from SND~\cite{SND:2020qmb, Achasov:2017vaq}.}
    \label{Fig:cs}
\end{figure*}

The decay widths predicted by our model are given in Table~\ref{tab:table4}. In an overall view, they are compatible with those of PDG within their errors.
\begin{table}[htb]
    \centering
    \renewcommand\arraystretch{1.5}
    \setlength{\tabcolsep}{14pt}
    \scalebox{0.8}{
    \begin{tabular}{c c c}
    \hline
    \hline
    Width                  &  This analysis                          &    PDG~\cite{ParticleDataGroup:2022pth}     \\
    \hline
    $\Gamma(K^\ast        \!\to\! K\pi)$          [$10^{-2}$ GeV]    &    $4.482\pm0.507$    &    $4.935\pm0.065$  \\
    $\Gamma(K^{\ast 0}    \!\to\! K^0\gamma)$     [$10^{-4}$ GeV]    &    $1.073\pm0.123$    &    $1.164\pm0.100$  \\
    $\Gamma(K^{\ast\pm}   \!\to\! K^\pm\gamma)$   [$10^{-5}$ GeV]    &    $4.290\pm1.085$    &    $5.040\pm0.470$  \\
    $\Gamma(a_2           \!\to\! \eta\pi)$       [$10^{-2}$ GeV]    &    $2.065\pm0.093$    &    $1.552\pm0.147$  \\
    $\Gamma(a_2           \!\to\! K\bar{K})$      [$10^{-3}$ GeV]    &    $5.680\pm0.462$    &    $5.243\pm0.890$  \\
    $\Gamma(a_2           \!\to\! \eta'\pi)$      [$10^{-4}$ GeV]    &    $5.103\pm1.186$    &    $5.885\pm1.002$  \\
    $\Gamma(a_2^\pm       \!\to\! \pi^\pm\gamma)$ [$10^{-4}$ GeV]    &    $6.270\pm2.412$    &    $3.114\pm0.323$  \\
    $\Gamma(K_2^\ast      \!\to\! K\pi)$          [$10^{-2}$ GeV]    &    $4.304\pm0.350$    &    $5.215\pm0.217$  \\
    $\Gamma(K_2^\ast      \!\to\! K^\ast\pi)$     [$10^{-2}$ GeV]    &    $2.197\pm0.585$    &    $2.581\pm0.180$  \\
    $\Gamma(K_2^\ast      \!\to\! \rho K)$        [$10^{-3}$ GeV]    &    $7.885\pm2.100$    &    $9.092\pm0.891$  \\
    $\Gamma(K_2^\ast      \!\to\! \omega K)$      [$10^{-3}$ GeV]    &    $2.293\pm0.611$    &    $3.031\pm0.842$  \\
    $\Gamma(K_2^{\ast\pm} \!\to\! K^\pm\gamma)$   [$10^{-4}$ GeV]    &    $3.361\pm1.376$    &    $2.400\pm0.503$  \\
    \hline
    \hline
    \end{tabular}}
    \caption{Fit to the decay widths of PDG~\cite{ParticleDataGroup:2022pth}. The cut-off is chosen as $\Lambda=2.5$ GeV.}
    \label{tab:table4}
\end{table}
Indeed, all the other decay widths are compatible with the data, except for the decay width of $\Gamma(a_2^\pm \!\to\! \pi^\pm \gamma)$, which deviates from the PDG about fifty percent. The reason may be that in this process, the final states could include a vector meson but not only pseudoscalars and/or photons. The vector meson will decay into the lighter mesons and/or photons, and one lacks the dynamical description of the FSI for these subsequent processes. Also, one would notice that ome other decay widths of the tensors, e.g., $\Gamma(a_2 \!\to\! \eta\pi)$, $\Gamma(K_2^\ast \!\to\! \omega K)$, and $\Gamma(K_2^{\ast\pm} \!\to\! K^\pm \gamma)$, have uncertainties about a quarter. They are all for the tensors, which have been inadequately studied until now. 
Nevertheless, the fit to the decay widths gives more constraints on the coupling constants and helps to extract the form factors more reliably. 
We apply the Bootstrap method~\cite{Efron:1979bxm} to obtain the uncertainties of our solution, which are calculated by varying the experimental data points within their errors and multiplying a normal distribution function. Notice that the uncertainties statistics from different cut-offs are included, too. 
In contrast, the errors from MINUIT~\cite{James:1975dr} are tiny and ignorable.

Our solutions fitting to the cross sections and invariant mass spectra are shown in Fig.~\ref{Fig:cs}. 
As observed, ours fits the data well. In the two left columns, the fits have different cut-offs. The solutions are represented as red dashed, blue dotted, black solid, purple dash-dotted, and green dash-dot-dotted lines with cut-offs of $\Lambda=2.3, 2.4, 2.5, 2.6, 2.7$ GeV, respectively. Specified, the lines with cut-offs $\Lambda=2.6, 2.7$ GeV exhibit poor performance compared with the others. See, e.g., the first graph at the bottom, where the results for the cross section of $e^+e^- \to K_S^0 K^{\pm} \pi^{\mp}$ is shown. The purple dash-dotted and green dash-dot-dotted lines, corresponding to the cut-offs $\Lambda=2.6, 2.7$ GeV, significantly deviate from the other lines in the energy range of [$1.4-1.6$] GeV, falling notably below the data points of BABAR. In contrast, the red dashed line, with cut-off $\Lambda=2.3$ GeV, looks slightly worse in the energy region above 2.2 GeV in the same graph. These are compatible with our strategy, where $\Lambda=2.5$ GeV is chosen as the optimal one.

The results with cut-off $\Lambda=2.5$ GeV are in the right two columns. The uncertainties of the solution, depicted as the cyan bands, are estimated from Bootstrap~\cite{Efron:1979bxm} method, as well as the cut-off dependence, where the two uncertainties are added together by the root mean square. The uncertainties adequately encompass most of the data points except for some invariant mass spectra such as $K_{S/L}^0 \pi^0$ and $K_{S}^0 K_{L}^0$ ones. This impacts our estimation of the anomalous magnetic moment of the muon, as discussed in the following section.

There is a broad peak in the energy region of $E_{\rm cm} \in [1.5-1.7]$ GeV for the cross sections. 
It may be caused by the complicated interaction involving the excited resonances, $V^{\prime(\prime\prime)}$, except for the $\phi(2170)$. 
The solution in this energy region is strongly constrained by the invariant mass spectra as given in the graphs in the right column of Fig.~\ref{Fig:cs}. 
Indeed, our solution fits the total cross sections better than that of the invariant mass spectra. In the second and fourth graphs in the second row of Fig.~\ref{Fig:cs}, the \lq peak' appears to have shifted slightly towards the left side. This shift may be attributed to the limitations in accurately describing the final state interactions between the pion and kaon, for which we have implemented an energy-dependent width for $K^\ast(892)$ to restore partly. See Eq.~(\ref{Eq:app:Gamma}). Additionally, the \lq peak' in the second and fourth graph of the third row appears to be slightly shifted to the right. This shift may be attributed to the absence of a precise model to describe $K\bar{K}$ FSI. 
All the experimental data sets of invariant mass spectra are from SND~\cite{SND:2020qmb, Achasov:2017vaq}, and the statistics are not high. Nevertheless, a combination fit to the invariant mass spectra and the rich data sets of the cross sections imposes a strong constraint on the parameters of our solution, which helps to give a reliable estimation of the $K\bar{K}\pi$ contribution to HVP.

\section{The muon anomalous magnetic moment}
With the cross sections obtained above, one can predict their contributions to the leading order (LO) HVP of the muon anomalous magnetic moment. One has~\cite{Brodsky:1967sr, Lautrup:1968tdb}
\begin{eqnarray}
    a_\mu^{\rm HVP,LO} = \frac{\alpha_e^2(0)}{3\pi^2} \int_{s_{\rm th}}^\infty ds \frac{\hat{K}(s)}{s} R_{\rm h}(s)~,
\end{eqnarray}
where $\alpha_e(0)=e^2/(4\pi)$ is the electromagnetic fine-structure constant and the kernel function $\hat{K}(s)$ can be found in Refs.~\cite{Brodsky:1967sr, Lautrup:1968tdb, Aoyama:2020ynm}. The hadronic $R$-ratio is derived as
\begin{eqnarray} \label{Eq:R}
    R_{\rm h}(s) = \frac{3s}{4\pi\alpha_e^2(s)} \sigma \bigg( e^+e^- \to {\rm hadrons} \bigg)~.
\end{eqnarray}
Here, the $\alpha_e(s)$ can be found in Refs.~\cite{Aoyama:2020ynm, Wang:2023njt}
\begin{eqnarray}
    \alpha_e(s) = \frac{\alpha_e(0)}{1-\Delta\alpha(s)}~, ~ ~ ~ \Delta\alpha(s) = \Pi'_\gamma(0)-\Pi'_\gamma(s)~, \nonumber
\end{eqnarray}
where $\Pi_\gamma(s)$ is the vacuum polarization operator. 
The total cross sections of $e^+e^- \to K\bar{K}\pi$ include three kinds of final states as~\cite{Keshavarzi:2018mgv}
\begin{eqnarray} \label{Eq:cs;all}
    \sigma(K\bar{K}\pi) \!=\! \sigma(K^+K^-\pi^0) \!+\! \sigma(K_S^0K_L^0\pi^0) \!+\! 2\sigma(K_S^0K^\pm\pi^\mp)\,. \nonumber
\end{eqnarray}
The prediction of the $(g-2)_\mu$ from $e^+e^- \to K\bar{K}\pi$ are given in Table~\ref{Tab:g-2} and their contributions to HVP from other works~\cite{Keshavarzi:2018mgv, Davier:2019can} are listed for comparison. The uncertainties are estimated from the Bootstrap method and cut-off dependence.
\begin{table}[htb]
    \centering
    \renewcommand\arraystretch{2.1}
    \setlength{\tabcolsep}{10pt}
    \scalebox{0.68}{
    \begin{tabular}{c c c c}
    \hline
    \hline
    $a_\mu^{\rm HVP,LO} \times 10^{10}$  &  This analysis  &  KNT18~\cite{Keshavarzi:2018mgv}  &  DHMZ19~\cite{Davier:2019can}  \\
    \hline
    $a_\mu^{K\bar{K}\pi}[1.260 \leq q \leq 1.937 ~ {\rm GeV}]$    &  $2.726\pm0.063$  &  $2.71\pm0.12$  &  $-$            \\
    $a_\mu^{K\bar{K}\pi}[{\rm th.} \leq q \leq 1.8 ~ {\rm GeV}]$  &  $2.407\pm0.062$  &  $2.44\pm0.11$  &  $2.45\pm0.13$  \\
    $a_\mu^{K\bar{K}\pi}[{\rm th.} \leq q \leq 2   ~ {\rm GeV}]$  &  $2.829\pm0.064$  &  $2.80\pm0.12$  &  $-$            \\
    $a_\mu^{K\bar{K}\pi}[{\rm th.} \leq q \leq 2.3 ~ {\rm GeV}]$  &  $3.178\pm0.071$  &  $-$            &  $-$            \\
    \hline
    \hline
\end{tabular}}
\caption{Our predictions of the contributions to the muon anomalous magnetic moment from $e^+e^- \to K\bar{K}\pi$. We also list the results of other works to give a comparison~\cite{Keshavarzi:2018mgv, Davier:2019can}. The ${\rm th.}$ denotes the threshold of $K\bar{K}\pi$.}
\label{Tab:g-2}
\end{table}
It is found that our estimation of the contributions of $K\bar{K}\pi$ is in good agreement with those gained by data-driven method~\cite{Keshavarzi:2018mgv, Davier:2019can} below 2 GeV. Moreover, our calculation of $K\bar{K}\pi$ channel contribution is up to 2.3 GeV within RChT, with the corresponding $a_{\mu}$ as $(3.178\pm0.071) \times 10^{-10}$. For reader's convenience, we also give an estimation of the $a_\mu$ from threshold up to 1.8 GeV, as $(2.407\pm0.062) \times 10^{-10}$. This is a bit smaller than those of Refs.~\cite{Keshavarzi:2018mgv, Davier:2019can}. It would be rather helpful if the experiments could perform measurements about the cross sections and angular distributions with higher statistics. 
Studies on other processes with multi-pseudoscalar, e.g., $e^+e^- \to 4\pi$, would refine the estimation of the theoretical prediction of HVP. In addition, further study on these processes would lead to building an efficient theory on low energy strong interactions and help to understand the discrepancy of $(g-2)_\mu$.

\section{Conclusion and summary}
In this work, we systematically studied the processes of $e^+e^- \to K^+K^-\pi^0$, $K_S^0K_L^0\pi^0$, and $K_S^0K^\pm\pi^\mp$ within the framework of resonance chiral theory. The experimental data of scattering cross sections, invariant mass spectra, and decay widths of the vectors and tensors are fitted to fix the unknown parameters. A high-quality solution is obtained. With it, we predict the LO HVP contributions to the muon anomalous magnetic moment, $a_\mu^{K\bar{K}\pi}=(3.178\pm0.071) \times 10^{-10}$, from threshold up to $E_{\rm cm}=2.3$ GeV, or for convenience of the reader, $a_\mu^{K\bar{K}\pi}=(2.407\pm0.062) \times 10^{-10}$, from threshold up to $E_{\rm cm}=1.8$ GeV. Further theoretical studies and experimental measurements on the electron-positron annihilation into hadrons are needed to improve the prediction on $(g-2)_\mu$ from the Standard Model.

\section*{Acknowledgements}
We thank the helpful discussions with Qin-He Yang, Di Guo, and Prof. Han-Qing Zheng. 
Especially, We are in debt to Prof. Jorge Portoles for his patient discussions all along. This work is supported by the National Natural Science Foundation of China with Grants No.12322502, 12335002, and U1932110, Hunan Provincial Natural Science Foundation with Grant No. 2024JJ3004, and Fundamental Research Funds for the central universities of China. Wen Qin is partly supported by the Hunan Provincial Department of Education with Grant No.22B0044 and the Hunan Provincial Natural Science Foundation with Grant No.2024JJ6300.

\appendix

\section{Feynman propagator and polarization of tensor} \label{App:1}
The Feynman propagator of the tensor is defined as~\cite{Bellucci:1994eb, Ecker:2007us}
\begin{eqnarray}
    G_{\mu\nu,\rho\sigma}^T(x) &=& \int \frac{d^4k}{(2\pi)^4} \frac{i P_{\mu\nu,\rho\sigma}(k)}{M_T^2-k^2-i\epsilon} {\rm e}^{-ikx}~,
\end{eqnarray}
where $k$ is the momentum, and one has
\begin{eqnarray}
    P_{\mu\nu,\rho\sigma}(k) &=& \frac{1}{2} (P_{\mu\rho} P_{\nu\sigma} + P_{\nu\rho} P_{\mu\sigma}) - \frac{1}{3} P_{\mu\nu} P_{\rho\sigma}~, \nonumber \\
    P_{\mu\nu} &=& g_{\mu\nu} - \frac{k_\mu k_\nu}{M_T^2}~. \nonumber
\end{eqnarray}
The tensor field operator $T_{\mu\nu}$ acting on the state of a spin-2 particle is expressed in terms of the polarization tensor $\varepsilon_{\mu\nu}(k,\lambda)$~\cite{Ecker:2007us}:
\begin{eqnarray}
    \langle 0 | T_{\mu\nu}(0) | T(k,\lambda) \rangle = \varepsilon_{\mu\nu}(k,\lambda)~.
\end{eqnarray}
with $\lambda$ the polarization. The sum over all polarizations gives
\begin{eqnarray}
    \sum_\lambda \varepsilon_{\mu\nu}(k,\lambda) \varepsilon_{\rho\sigma}^\ast(k,\lambda) = P_{\mu\nu,\rho\sigma}(k)~. \nonumber
\end{eqnarray}

\section{Effective Lagrangians with tensor} \label{App:2}
As is known, the effective Lagrangians should satisfy discrete symmetries~\cite{Ecker:1988te, Bijnens:1999sh, Fettes:2000gb}. The properties of chiral operators transforming under the parity ($P$), charge conjugation ($C$), and hermiticity (h.c.) are given in Table~\ref{tab:operator}.
\begin{table}[htb]
\centering
\vspace{0.5cm}
\renewcommand\arraystretch{1.4}
\setlength{\tabcolsep}{12pt}
\scalebox{0.93}{
\begin{tabular}{c c c c c}
\hline
\hline
Operator            &    Dim          &    $P$                     &    $C$                         &    h.c.              \\
\hline
$u_\mu$             &    $1$          &    $-u^\mu$                &    $(u_\mu)^T$                 &    $u_\mu$           \\
$h_{\mu\nu}$        &    $2$          &    $-h^{\mu\nu}$           &    $(h_{\mu\nu})^T$            &    $h_{\mu\nu}$      \\
$\chi_\pm$          &    $2$          &    $\pm \chi_\pm$          &    $(\chi_\pm)^T$              &    $\pm \chi_\pm$    \\
$f^{\mu\nu}_\pm$    &    $2$          &    $\pm f_{\pm \mu\nu}$    &    $\mp (f^{\mu\nu}_\pm)^T$    &    $f^{\mu\nu}_\pm$  \\
$V_{\mu\nu}$        &    $0$          &    $V^{\mu\nu}$            &    $-(V_{\mu\nu})^T$           &    $V_{\mu\nu}$      \\
$T_{\mu\nu}$        &    $0$          &    $T^{\mu\nu}$            &    $(T_{\mu\nu})^T$            &    $T_{\mu\nu}$      \\
$\varepsilon_{\mu\nu\rho\sigma}$ & $0$ & $-\varepsilon^{\mu\nu\rho\sigma}$ & $\varepsilon_{\mu\nu\rho\sigma}$ & $\varepsilon_{\mu\nu\rho\sigma}$  \\
\hline
\hline
\end{tabular}}
\caption{The chiral dimension (Dim), $P$, $C$ and h.c. transformation properties of operators for constructing chiral Lagrangians.}
\label{tab:operator}
\end{table}
Following it, we construct the Lagrangians about TJP and TVP terms mentioned in the above sections. 
We use the following constraints to select the linearly independent terms among all the possible combinations of the operators~\cite{Ruiz-Femenia:2003jdx}:

\noindent (i) Equations of motion (EOM)~\cite{Bijnens:1999sh}.
\begin{equation}
\nabla_\mu u^\mu = \frac{i}{2} \left( \chi_- - \frac{1}{n_f} \langle \chi_- \rangle \right)~,
\end{equation}
with $n_f$ the number of light flavors ($n_f=3$ in our case). With this equation, $\nabla_\mu u^\mu$ will not appear in the chiral effective Lagrangians as it can be replaced by $\chi_-$.

\noindent (ii) Total derivative~\cite{Fearing:1994ga, Ebertshauser:2001nj}.
\begin{eqnarray}
    \langle \nabla_\mu \left( A B C \cdots \right) \rangle &=& \langle \left( \nabla_\mu A \right) B C \cdots \rangle + \langle A \left( \nabla_\mu B \right) C \cdots \rangle \nonumber \\[2mm]
    && + ~ \langle A B \left( \nabla_\mu C \right) \cdots \rangle + \cdots~,
\end{eqnarray}
where $\nabla_\mu$ is the covariant derivative, and $A, B, C, \cdots$ represent the operators. The total derivative would lead to a vanished action integrated from the Lagrangians with a total derivative. Correspondingly, one should reduce one of the terms on the right side of the equal sign as they are not independent.

\noindent (iii) Schouten identity.
\begin{eqnarray}
g_{\alpha\lambda}\varepsilon_{\mu\nu\rho\sigma} + g_{\alpha\mu}\varepsilon_{\nu\rho\sigma\lambda} + g_{\alpha\nu}\varepsilon_{\rho\sigma\lambda\mu} \nonumber \\
+ g_{\alpha\rho}\varepsilon_{\sigma\lambda\mu\nu} + g_{\alpha\sigma}\varepsilon_{\lambda\mu\nu\rho} = 0~.
\end{eqnarray}
With these constraints, we are able to pick out the independent effective Lagrangians, and finally, the TJP and TVP interaction Lagrangians are given in Eqs.~(\ref{Eq:TJP}, \ref{Eq:TVP}).

\section{Notations for the form factors} \label{App:3}
The notations of the form factors employed in the text are specified below:
\begin{widetext}
\begin{eqnarray}
    GR_1(q^2,s) &=& (g_1+2g_2-g_3)(q^2+s-4m_K^2-m_\pi^2) + 4(2g_4+g_5)m_K^2 - 4g_2(q^2-2m_K^2-m_\pi^2) - 4g_4(m_K^2-m_\pi^2)~, \nonumber \\
    GR_2(q^2,s) &=& (g_1+2g_2-g_3)(q^2-s-m_\pi^2) + 2(2g_4+g_5)m_\pi^2 - 2g_2(q^2-2m_K^2-m_\pi^2) + 4g_4(m_K^2-m_\pi^2)~, \nonumber \\
    CR_1(q^2,x,m^2) &=& (c_1-c_2+c_5)q^2 - ( c_1-c_2-c_5+2c_6)x + (c_1+c_2+8c_3-c_5)m^2~, \nonumber \\
    CR_2(q^2,x) &=& (c_1-c_2+c_5)q^2 - (c_1-c_2-c_5+2c_6)x + (c_1+c_2+8c_3-c_5)m_K^2 + 24c_4(m_K^2-m_\pi^2)~, \nonumber \\
    DR(q^2,x,m^2) &=& (d_1+8d_2-d_3)m^2 + d_3(q^2+x)~, \nonumber \\
    CR^T(q^2,s,x) &=& (2c_1^T-2c_2^T-c_3^T) \Big[ M_{K_2^\ast}^2(q^2-2s-x+3m_K^2) + (q^2-x-m_K^2)(m_K^2-m_\pi^2) \Big] \nonumber \\
    && +2c_3^T (M_{K_2^\ast}^2-x)(m_K^2-m_\pi^2)~, \nonumber \\
    DR^T(q^2,s,x) &=& (2d_1^T-2d_2^T-d_3^T) \Big[ M_{K_2^\ast}^2(q^2-2s-x+3m_K^2) + (q^2-x-m_K^2)(m_K^2-m_\pi^2) \Big] \nonumber \\
    && +2d_3^T (M_{K_2^\ast}^2-x)(m_K^2-m_\pi^2)~. \label{Eq:CRDR}
\end{eqnarray}
\end{widetext}

\section{The energy-dependent widths of the vector resonances} \label{App:4}
The energy-dependent widths of the vector resonances taken from Refs.~\cite{Dai:2013joa, Dumm:2009va, Jamin:2006tk}
\begin{eqnarray} \label{Eq:app:Gamma}
    \Gamma\!_\rho(s)\! &=& \! \frac{M_\rho s}{96\pi F^2} \Big[ \sigma_\pi^3(s) \, \theta(s-4m_\pi^2) \nonumber \\
    && +\frac{1}{2} \sigma_K^3(s) \, \theta(s-4m_K^2) \Big]~, \nonumber \\
    \Gamma\!_{\rho'}(s)\!\! &=& \!\!\Gamma\!_{\rho'} \frac{\sqrt{s}}{M_{\rho'}} \Big[ \frac{\sigma_\pi^3(s)}{\sigma_\pi^3(M_{\rho'}^2)} \Big] \theta(s-4m_\pi^2)~, \nonumber \\
    \Gamma\!_{\rho''}(s) \!\! &=& \!\! \Gamma\!_{\rho''} \frac{\sqrt{s}}{M_{\rho''}} \Big[ \frac{\sigma_\pi^3(s)}{\sigma_\pi^3(M_{\rho''}^2)} \Big] \theta(s-4m_\pi^2)~, \nonumber \\
    \Gamma\!_{K^{\ast}}\!(s)\!\! &=& \!\! \frac{M_{K^{\ast}}}{128\pi F^2 s^2} \Big[ \lambda^{\frac{3}{2}}(s,m_K^2,m_\pi^2) \, \theta[s\!-\!(m_K\!+\!m_\pi)^2] \nonumber \\
    && +\lambda^{\frac{3}{2}}(s,m_K^2,m_\eta^2) \, \theta[s-(m_K+m_\eta)^2] \Big]~, \nonumber \\
    \Gamma\!_{K^{\ast'}}\!(s)\!\! &=& \!\! \Gamma\!_{K^{\ast'}} \! \frac{\! M_{K^{\ast'}}^4}{s^2} \! \Big[\! \frac{\lambda(\!s,\!m_K^2,\!m_\pi^2)}{\lambda(\!M_{K^{\ast'}}^2,\!m_K^2,\!m_\pi^2)} \!\Big]^{\frac{3}{2}} \! \theta[s\!-\!(\!m_K\!+\!m_\pi\!)^2]~, \nonumber \\
    \Gamma\!_{K^{\ast''}}\!(s)\!\! &=& \!\! \Gamma\!_{K^{\ast''}} \! \frac{\! M_{K^{\ast''}}^4}{s^2} \! \Big[\! \frac{\lambda(\!s,\!m_K^2,\!m_\pi^2)}{\lambda(\!M_{K^{\ast''}}^2,\!m_K^2,\!m_\pi^2)} \!\Big]^{\frac{3}{2}} \! \theta[s\!-\!(\!m_K\!+\!m_\pi\!)^2]~. \nonumber \\[3mm]
\end{eqnarray}
where $\sigma_P(s)=\sqrt{1-4m_P^2/s}$ is the phase space factor and $\theta(x)$ is the step function.

\section{Two-body decays} \label{App:5}
The two-body decay widths of vectors and tensors are listed below,
\begin{widetext}
\begin{eqnarray}
    \Gamma(K^\ast \to K\pi) &=& \frac{G_V^2}{64\pi F^4} \frac{\lambda^{3/2}(M_{K^\ast}^2,m_K^2,m_\pi^2)}{M_{K^\ast}^3}, \nonumber \\
    \Gamma(K^{\ast 0} \to K^0\gamma) &=& \frac{\alpha(M_{K^\ast}^2\!-\!m_K^2)^3}{24M_{K^\ast}^5} \bigg\{ \frac{4\sqrt{2}}{3F M_V} \big[ (c_1\!+\!c_2\!+\!8c_3\!-\!c_5)m_K^2 \!-\! (c_1\!-\!c_2\!-\!c_5\!+\!2c_6)M_{K^\ast}^2 \big] \nonumber \\
    && -\frac{2F_V}{F} \bigg( \frac{1}{M_\rho^2} \!-\! \frac{1}{3M_\omega^2} \!+\! \frac{2}{3M_\phi^2} \bigg) \big[ (d_1\!+\!8d_2\!-\!d_3)m_K^2 \!+\! d_3M_{K^\ast}^2 \big] \bigg\}^2, \nonumber \\
    \Gamma(K^{\ast\pm} \to K^\pm\gamma) &=& \frac{\alpha(M_{K^\ast}^2\!-\!m_K^2)^3}{24M_{K^\ast}^5} \bigg\{ \frac{2\sqrt{2}}{3F M_V} \big[ (c_1\!+\!c_2\!+\!8c_3\!-\!c_5)m_K^2 \!-\! (c_1\!-\!c_2\!-\!c_5\!+\!2c_6)M_{K^\ast}^2 \!+\! 24c_4(m_K^2\!-\!m_\pi^2) \big] \nonumber \\
    && -\frac{2F_V}{F} \bigg( \frac{1}{M_\rho^2} \!+\! \frac{1}{3M_\omega^2} \!-\! \frac{2}{3M_\phi^2} \bigg) \bigg[ (d_1\!+\!8d_2\!-\!d_3)m_K^2 \!+\! d_3M_{K^\ast}^2 \bigg] \bigg\}^2, \nonumber \\
    \Gamma(a_2 \to \eta\pi) &=& \frac{g_T^2}{360\pi F^4} \frac{\lambda^{5/2}(M_{a_2}^2,m_\eta^2,m_\pi^2)}{M_{a_2}^7} \bigg\{ -2\sqrt{2}\sin(2\theta_P)-\cos(2\theta_P)+3 \bigg\}, \nonumber \\
    \Gamma(a_2 \to K\bar{K}) &=& \frac{g_T^2}{120\pi F^4} \frac{(M_{a_2}^2-4m_K^2)^{5/2}}{M_{a_2}^2}, \nonumber \\
    \Gamma(a_2 \to \eta'\pi) &=& \frac{g_T^2}{360\pi F^4} \frac{\lambda^{5/2}(M_{a_2}^2,m_{\eta'}^2,m_\pi^2)}{M_{a_2}^7} \bigg\{ 2\sqrt{2}\sin(2\theta_P)+\cos(2\theta_P)+3 \bigg\}, \nonumber \\
    \Gamma(a_2^\pm \to \pi^\pm\gamma) &=& \frac{\alpha}{80} \bigg( \frac{M_{a_2}^2-m_\pi^2}{M_{a_2}} \bigg)^5 \bigg\{ -\frac{\sqrt{2}}{F}(2c_1^T-2c_2^T-c_3^T) + \frac{F_V}{F M_\rho^2}(2d_1^T-2d_2^T-d_3^T) \bigg\}^2, \nonumber \\
    \Gamma(K^\ast_2 \to K\pi) &=& \frac{g_T^2}{80\pi F^4} \frac{\lambda^{5/2}(M_{K^\ast_2}^2,m_K^2,m_\pi^2)}{M_{K^\ast_2}^7}, \nonumber \\
    \Gamma(K^\ast_2 \to K^\ast\pi) &=& \frac{3(2d_1^T-2d_2^T-d_3^T)^2}{1280\pi F^2 M_{K^\ast}^2} \frac{\lambda^{5/2}(M_{K^\ast_2}^2,M_{K^\ast}^2,m_\pi^2)}{M_{K^\ast_2}^5}, \nonumber \\
    \Gamma(K^\ast_2 \to \rho K) &=& \frac{3(2d_1^T-2d_2^T-d_3^T)^2}{1280\pi F^2 M_\rho^2} \frac{\lambda^{5/2}(M_{K^\ast_2}^2,M_\rho^2,m_K^2)}{M_{K^\ast_2}^5}, \nonumber \\
    \Gamma(K^\ast_2 \to \omega K) &=& \frac{(2d_1^T-2d_2^T-d_3^T)^2}{1280\pi F^2 M_\omega^2} \frac{\lambda^{5/2}(M_{K^\ast_2}^2,M_\omega^2,m_K^2)}{M_{K^\ast_2}^5}, \nonumber \\
    \Gamma(K^{\ast\pm}_2 \to K^\pm\gamma) &=& \frac{\alpha}{80} \bigg( \frac{M_{K_2^\ast}^2\!-\!m_K^2}{M_{K_2^\ast}} \bigg)^5 \bigg\{ -\frac{\sqrt{2}}{F}(2c_1^T\!-\!2c_2^T\!-\!c_3^T) \!+\! \frac{F_V}{2F} \bigg( \frac{1}{M_\rho^2} \!+\! \frac{1}{3M_\omega^2} \!+\! \frac{2}{3M_\phi^2} \bigg)(2d_1^T\!-\!2d_2^T\!-\!d_3^T) \bigg\}^2. \nonumber \\
\end{eqnarray}
\end{widetext}

\bibliography{ref}

\end{document}